\documentstyle[a4,12pt]{article}
\begin{document}
\input{epsf}
\bigskip
\bigskip
\hfil{SWAT/227}
\bigskip
\bigskip
\centerline{\bf The Dirac Sea Contribution To The Energy  Of An Electroweak String}
\bigskip
\bigskip
\centerline{\bf Martin Groves and Warren B. Perkins,}
\bigskip

\centerline{\it Department Of Physics,}
\centerline{\it University of Wales Swansea,}
\centerline{\it Singleton Park,}
\centerline{\it Swansea, SA2 8PP,}
\centerline{\it Great Britain.}

\bigskip
\bigskip
\bigskip
\bigskip
\centerline{\bf Abstract}
\bigskip
We present a systematic determination of the order $\hbar$ fermionic
energy shift when an electroweak string is perturbed. We show that
the combined effect of zero modes, bound states and continuum
states is to lower the total fermionic ground state energy of the string 
when the Higgs instability of the string is excited.  The effect of the
Dirac sea is thus to destabilise the string. However, this effect
can be offset by populating positive energy states. Fermions enhance
the stability of an electroweak string with sufficiently populated fermionic
bound states.  

\vfil\eject
\noindent{\bf 1. Introduction}

\bigskip
The resurrection of interest in electroweak strings, first 
discovered by Nambu \cite{Nambu}, induced by the discovery of the W and 
Z-string solutions \cite{Vas1} lead to  intense research 
into the properties and potential impact of 
these objects. Both the W and Z-strings solutions are U(1) Nielsen-Olesen 
vortices embedded in some U(1) subgroup of the Weinberg-Salam 
model \cite{embed}. Unlike the archetypal Abelian Higgs Nielsen-Olesen vortex 
and standard GUT vortices, the stability of these embedded strings is not 
guaranteed by any topological argument. The enlarged field 
 space of the SU(2)xU(1) Salam Weinberg model provides 
extra directions in field space which allow the 
construction of continuous, finite energy paths that unwind the string 
solution.  
The criterion for the existence of 
topological strings is that the first homotopy group of the vacuum 
manifold is non-trivial \cite{homot}. The first homotopy group of the three sphere,
which is the vacuum manifold of the electroweak theory,  
is however trivial and therefore the electroweak string  solutions are 
non-topological in nature. This implies that the W and Z-strings are 
just finite energy solutions to the classical field equations and, 
at best, electroweak stings could sit at a local 
minimum of the electroweak energy functional. They would then 
be separated from, and prevented from  decaying  into,  the trivial vacuum by a finite 
energy barrier. This barrier would either have to be surmounted or 
quantum mechanically tunnelled through  to destroy the string solution. 
Alternatively electroweak strings could be local maxima or saddle 
points of the energy functional, in which case the pure string 
solutions would be intrinsically unstable. The issue of stability is a 
question of energetics and depends on the value of the parameters used. 
Detailed stability analyses of the Z-string have been carried out by two 
independent sets of authors\cite{James,Good}. Both investigations 
concluded that the Z-string solution possess modes of instability 
except in the unphysical region where $sin^2(\theta_W)\sim  1$. Other investigations\cite{winstab}
 imply that the W-string solutions 
are unstable for all values of the electroweak parameters. W-strings 
will not be considered any further during the course of this paper.

The instability of electroweak strings for physical values of the 
Standard Model parameters would tend to imply that electroweak strings are 
physically insignificant. Had electroweak strings been stable then we 
would have expected  a network of strings to be  formed during 
the electroweak phase transition through the Kibble mechanism. The 
density of electroweak strings would be lower than their topological 
counterparts as  the 
vanishing of  the electroweak Higgs field at a point does not necessarily imply the 
existence of a string. Another important difference arises because 
electroweak strings would tend to form fragments of finite length that are 
bounded by a monopole anti-monopole pair.  As discussed in \cite{Annebg}
the inherent 
contraction of these string fragments within the electroweak theory 
provide all of the conditions required for electroweak baryogensis. 
Although the CP violation in the minimal electroweak theory is too 
small to account for the observed baryon anti-baryon asymmetry, 
electroweak string are also expected to arise in   extensions of 
the minimal electroweak which admit greater CP violation \cite{twohiggs}.

The stability analysis discussed thus far has focussed on the pure 
Z-string solutions. It is however possible that the formation 
of bound states of other particles on the string may have a non-trivial 
effect on the stability of electroweak strings. Of particular 
importance for a string like solution is the effect of 
a finite density of bound states moving along the 
string.  In \cite{Wat} the authors 
considered a toy model in which they extended the electroweak theory by 
adding an extra global U(1) scalar to the bosonic sector of the 
electroweak theory. They constructed solutions with a finite U(1) 
charge per unit length of the string  and then investigated the 
stability of the resulting solutions. They found that it was 
possible  to extend the region in which the electroweak string is stable 
 to $sin^2(\theta_W) \sim 0.47$. Thus the stability of an electroweak 
string may be enhanced by its interaction with other fields. The stabilising
effects of a plasma
of scalar particles have also been considered \cite{Nagasawa}.
 
 The authors of \cite{Wat} rightly point out that, 
in the context of the electroweak theory, the pertinent particles to 
consider are the fermions which gain their mass via a Yukawa coupling 
to the Higgs field. As will be discussed in this paper, the 
systematic analysis of  fermionic states in the  electroweak string background 
is a far more involved task than the analysis of bosonic condensates. 
The critical difference arises because of the different statistics 
satisfied by fermions and bosons. Bose-Eienstein statistics allow a 
significant population of the bosonic ground state and hence a 
classical treatment of this single mode is appropriate. In sharp contrast, the Pauli 
exclusion principle does not allow a similiar classical treatment in the
fermionic case and a complete 
field theoretic analysis is required. Naively, as the fermions gain 
their mass through a Yukawa coupling to the Higgs field it would be 
expected that the massive fermions would rather be localised about the 
centre of the string and prevent the string from decaying in an attempt 
to maintain their low energy state. The purpose of this paper is to 
carryout a systematic analysis of fermions in the background of the Z-string. 
The effects of the fermionic zero mode on the stability of the string was 
considered in \cite{nac,hong} where it is claimed that the fermionic ground 
state energy makes a negative contribution to the energy of the string 
and increases the instability. However, we must consider the full Dirac 
spectrum.  It is 
not clear  what effect the infinite number of continuum 
states have upon  string stability and positive 
energy bound states can only be consistently included if the actual 
numerical value of the fermionic ground state energy is calculated. 
Here we present a 
systematic, quantitaive treatment of the fermionic contribution to the 
energy of the Z-string.

It will be shown that a systematic treatment of fermions in a  string 
background requires a complete field theoretic calculation. 
Techniques are developed for the computation and  renormalisation 
of the Dirac sea contribution to the energy of the Z-string. Many of 
the techniques developed have a wider range of applicability,  particularly
for determining Dirac sea energies in other 3+1D backgrounds such as the 
sphaleron. Having 
shown how the Dirac sea energy can be computed in the Z-string 
background,  the change in this energy is computed when a mode 
of instability of the Z-string is excited. This determines the contribution
of the fermionic ground state to  the stability of the
Z-string. Having understood the ground state energy it is 
possible to consistently consider the effects of filled, positive energy 
bound states on the stability of the string. 

In section 2 we review the electroweak string and its stability.
A systematic discussion of the semi-classical treatment of 
fermion-soliton interactions is given in section 3. In this section the origins of the Dirac 
sea contribution to the fermionic energy of a soliton are demonstrated 
from a field theoretic point of view and a formal statement of the 
required calculation is made. 
In section 4  we review the Dirac equation in 
the background of the Z-string and derive the whole fermionic 
spectrum. 

The fermionic ground state energy is divergent and 
its computation requires the use of  a regulator and renormalisation 
scheme. In section 5  the proper time regularisation procedure is 
applied and an expression for the regulated  fermionic energy of the pure 
Z-string is derived.  
We present a consistent renormalisation of the regularised fermionic energy
in section 7. The appropriate counter terms are derived using
the heat kernel expansion.

In order to calculate the fermionic energy, we must sum over all continuum states.
This sum is implemented by discretising the continuum, determining momentum shift
functions and analytically recovering the continuum. The formalism for this procedure
is established in section 6.  To demonstrate the consistency of this procedure and
highlight its key physical properties, we analyse a core model string in section 8.
The core model 
introduces simplified string profile functions which allow the Dirac 
equation to be solved analytically and allows significant insight into 
numerous subtle aspects of the calculation of the fermionic energy to 
be gained.  

The calculation of real physical interest is presented in section 9.
We introduce the perturbed string, determine the Dirac spectrum in
this background and evaluate the counter terms.
We then apply the formalism of sections 5 to 7 to determine the 
difference between the ground state fermionic energy of the perturbed string
and that of the pure Z-string.

Once the change in the fermionic ground state energy has been 
calculated it is only a small step to add a finite density of positive 
energy bound states moving along  the string. In section 9 we 
investigate the effect of changing the density of positive energy states 
upon the stability of the string and discuss the possibility of stabilising the
electroweak string.

We conclude with a summary of what has been learned during 
this investigation and point towards future avenues of research.

\bigskip
\bigskip
\noindent{\bf 2. Strings in the Electroweak Theory}

\bigskip

In this section we review the Z-string solution and its stability.
The existence of string in the electroweak theory was first proposed by 
Nambu \cite{Nambu} in 1977. Interest in electroweak string was however 
resurrected by Vachaspati \cite{Vas1} with the discovery 
of the Z-string and W-string 
solutions.

To set notation, we take the purely bosonic sector of the electroweak theory to
be  defined by the Lagrangian density,
\begin{equation}
L_{Bosonic} = -\frac{1}{4} W_{\mu \nu}^a W^{a \mu \nu } 
-\frac{1}{4} F_{\mu \nu} F^{\mu \nu} 
+\mid D_\mu \Phi \mid^2 - \lambda (\mid \Phi\mid^2 - \eta^2)^2\label{int1}
\end{equation}
where,
\begin{equation}
 W_{\mu \nu}^a = \partial_\mu W_\nu^a - \partial_\nu W_\mu^a 
+ \hat g \epsilon^{abc} W_\mu^b W_\nu^c,\label{int2}
\end{equation}
\begin{equation}
F_{\mu \nu} = \partial_\mu B_\nu - \partial_\nu B_\mu,\label{int3}
\end{equation}
\begin{equation}
D_\mu \Phi = (\partial_\mu - \frac{i \hat g }{2} \sigma^a W^a_\mu 
-\frac{i g^\prime}{2} B_\mu ) \Phi, \label{int4}
\end{equation} 
and,
\begin{equation}
\Phi = \left( \begin{array}{c}
\phi_+ \\ \phi_o 
\end{array}
\right), \label{int4.5}
\end{equation}
is the standard Higgs doublet.
The gauge boson mass eigenstates are defined in the usual way,  
\begin{equation}
\left( \begin{array}{c}
Z_\mu \\ A_\mu 
\end{array} \right) = 
\left( \begin{array}{cc}
\cos \theta_w & - \sin\theta_w \\
\sin \theta_w & \cos\theta_w
\end{array}
\right)
\left(\begin{array}{c}
W_\mu^3 \\ B_\mu 
\end{array}
\right),\quad \label{int5}
W^\pm_\mu = \frac{1}{\sqrt{2}}(W^1_\mu \pm i W^2_\mu), \label{int6}
\end{equation}
where $ \theta_w $ is the Weinberg angle, 
$\hat g  = 2 q \cos \theta_w$ and
$g^\prime  =  2 q \sin \theta_w$.

The Z-string ansatz is defined by \cite{Vas1}, 
\begin{equation}
Z_1 =   -\frac{\nu(r)}{q r } \sin \theta ,\quad
Z_2 =  \frac{\nu(r)}{q r } \cos \theta ,\quad
\phi_o =  \eta f(r) e^{i \theta}, \label{int8}
\end{equation}
with all other fields in the theory set to zero. By keeping only the fields which are non-zero
 in the Z-string background, the energy of the field configuration becomes, 
\begin{equation}
E= \int d^3x \left( \frac{1}{4} Z_{ij}^2 + 
\mid \partial_i \phi_o -i q Z_i \phi_o \mid^2
+\lambda (\mid \phi_o \mid^2 - \eta^2)^2 \right), \label{int9}
\end{equation}
where the indices $i, j = 1,2,3$.
Rescaling the 
radial variable by $ r \rightarrow M_z r $, allows the energy per unit 
length of the string (in units of $M_z^2$) to be written as, 
\begin{equation}
E_{string} = 2\pi \int_0^\infty dr r \left( 
\left( \frac{\nu^\prime}{r} \right)^2 + (f^\prime)^2 
+\left( \frac{f}{r} \right)^2 (\nu-1)^2 
+\frac{1}{4} \beta^2 (f^2-1)^2 \right),\label{int10}
\end{equation}
where,
$
 \beta=\frac{M_H}{M_Z}. 
$ 
Variation of (\ref{int10}) with respect to the profile functions 
$f(r)$ and $\nu(r)$,  leads to the familiar Nielsen-Olesen equations \cite{novort} for a unit 
winding number string,
\begin{equation}
f^{\prime \prime}(r)+\frac{f^{\prime}(r)}{r}-\frac{f(r)}{r^2}(\nu(r)-1)^2
-\beta (f(r)^2-1)f(r) =0, \label{int11}
\end{equation}
\begin{equation}
\nu^{\prime \prime}(r)-\frac{\nu^\prime (r)}{r}-(\nu(r)-1)f^2(r)=0. 
\label{int12}
\end{equation}

Thus the Z-string is a $ U(1) $ Nielsen-Olesen string embedded 
in the electroweak theory in the direction of the $U(1)_Z$ subgroup 
generated by the $Z$ gauge particle. The Z-string is not
topologically stable, such a field configuration can decay to the trivial
vacuum, and stability is an issue of energetics.
By searching the $(\beta, \sin^2 \theta_w) $ plane it is found that the 
electroweak string in unstable except for $\sin^2 \theta_w \approx 1 $ \cite{James,Good}. 
Of particular importance,  electroweak strings are unstable for the 
physically observed  
values of the electroweak parameters ( $\sin^2 \theta_w \approx 0.23).$ 

For a unit winding number string  two modes of instability were found. The first  
corresponding to the upper component of the Higgs doublet, $\phi_+$, acquiring 
a non-zero value in the string core and the second 
corresponding to the W field,
\begin{equation}
W_{\downarrow}^- = \frac{e^{i \theta}}{\sqrt{2}} (
W_r^- + \frac{i}{r} W_\theta^- ),
\end{equation}
forming a non-zero value inside the string core. The latter instability 
corresponds to the formation of a W-condensate inside the string, this mode 
of instability has been investigated in \cite{perk1,jog}.
 The effect of these 
modes is to cause the string to unwind and decay to the trivial vacuum. 


The stability of electroweak string may however be changed by the 
presence of  bound states. Vachaspati and Watkins \cite{Wat} found that, by coupling the 
bosonic sector of the electroweak theory to a global complex 
scalar field, it is possible to significantly enhance the 
stability of the Z-string solution by the formation of scalar bound states. 
However,  as they point out,  the pertinent particles to consider are the fermions
of the standard model. As
the Higgs field vanishes in the core of the string, fermions 
which gain their mass from a Yukawa coupling to the Higgs field will be 
massless inside the string core. 
Thus naively they would rather reside within the core of the string and resist its 
dissolution. 

While the scalar case can be considered classically,
a  purely classical analysis is not possible for fermions 
as their statistics prevent a 
large occupation number within a given energy state. Consequently
a full quantum mechanical treatment is required.

Fermions in the background of the electroweak string have been considered by 
several authors \cite{boss,nac,hong}. However, only the effects of
the zero mode states were considered.
We present a systematic calculation of the change 
in the entire fermionic energy brought about by perturbing the 
Z-string.

\bigskip
\bigskip
\noindent{\bf 3. Fermion-Soliton Interactions and the Dirac Sea} 

\bigskip

The fermionic part of the electroweak action can 
be written in the form,
\begin{equation}
i \int d^4x  \Psi^\dagger \hat{A} [\Phi,Z_\mu] \Psi \label{Ds1}
\end{equation}
where $\hat{A} [\Phi,Z_\mu]$ is some operator which depends upon the 
background scalar and gauge fields and the fermionic fields are grouped 
together in the spinor $\Psi$.
The fermionic contribution to the path integral
is a Gaussian functional 
integral over the fermionic fields. 
For anti-commuting fields this Gaussian 
functional integral has the standard result \cite{raj},
\begin{equation}
\int D[\Psi^\dagger(x,t)] D[\Psi (x,t)] \exp \left(
i \int d^4x \Psi^\dagger \hat{A} [\Phi,Z_\mu] \Psi \right) 
= \mbox{Det } \hat{A} [\Phi,Z_\mu]. \label{Ds2}
\end{equation}
Using (\ref{Ds2}) it is possible to formally integrate out the fermionic 
contribution to the action and obtain  an effective action which 
encapsulates the effects of the fermion fields yet only contains explicit 
dependence on the bosonic fields.  This can rarely be done in practice 
because the calculation of the determinant in  
(\ref{Ds2}) requires a knowledge of all 
the eigenvalues of the operator $\hat{A} [\Phi,Z_\mu]$ for a general 
background. 
Instead we consider specific backgrounds.

The operator 
$\hat{A} [\Phi,Z_\mu]$ is related to the Dirac Hamiltonian 
$\hat{H}[\Phi,Z_\mu]$  in the bosonic 
background of the fields $\Phi$ and $Z_\mu$ by,
\begin{equation}
\hat{A} [\Phi,Z_\mu] =  i \frac{\partial}{\partial t} - \hat{H}[\Phi,Z_\mu].
\end{equation}
For a static field configuration the Dirac Hamiltonian has the set of 
energy eigenvalues $\{ E_r \}$ which satisfy,
\begin{equation}
\hat{H}[\Phi,Z_\mu] \Psi_r = E_r \Psi_r. \label{Ds3.1}
\end{equation}
This is just the Dirac 
equation describing the fermions in the background of the string.

The determinant in (\ref{Ds2}) can be expressed 
in terms of the energy eigenvalues  \cite{raj},
\begin{equation}
\mbox{Det } \hat{A} [\Phi,Z_\mu] = \sum_{\{ n_r \}} C (\{n_r\}) 
\exp \left[ -i T \left( \sum_r (-E_r + n_r E_r) \right) \right] \label{Ds3.2}
\end{equation}
where the integers $n_r$ 
are the occupation numbers of the excited positive energy states
and $ C ( \{ n_r \}) $ are the associated 
combinatoric degeneracy factors.

The fermionic ground state energy is then given by,
\begin{equation}
E_{gs} = \sum_{r} (-E_r). \label{Ds3.3}
\end{equation}
This is exactly the energy which would arise 
from filling all the negative energy states as predicted by Dirac.

The ground state energy of the field configuration is then given 
by, 
\begin{equation}
E_{ground} = E_{Classical} + \left( E_{bos} + \sum_r (- E_r ) 
\right) \label{Ds4}
\end{equation}
where $E_{bos}$ is the effect of the bosonic fluctuations (these 
will not be considered here).

We will be interested in  the difference between the Dirac sea energies
 of two systems, this will generally be referred to as the 
fermionic energy difference, which can be written as, 
\begin{equation}
\delta E_{fermion} = \sum_{r < 0 } \left( E_{r,1} - E_{r,2} \right),
\label{Ds5}
\end{equation}
where $ E_{r,1}$ and $ E_{r,2}$ are the energy eigenvalues of the Dirac 
operator in the two backgrounds and the sum is over all 
the negative energy states. It is useful to 
express the fermionic energy difference (\ref{Ds5}) as a 
functional trace over a complete set of eigenfunctions. Specifically,
\begin{eqnarray}
\delta E_{fermion} & =& -\frac{1}{2} \mbox{Tr }\left( \mid \hat{H}_1 \mid - 
\mid \hat{H}_2 \mid \right) 
= -\frac{1}{2}  \mbox{Tr }\left( \sqrt{\hat{H}_1^2} - 
\sqrt{\hat{H}_2^2} \right), \label{Ds6}
\end{eqnarray}
where $\hat{H}_{1,2}$ are the Dirac Hamiltonians for the two systems and
the factor of $\frac{1}{2}$ arises because the trace in (\ref{Ds6}) is over 
both the positive and negative energy eigenstates.  These are known to be 
in a $1:1$ correspondence through the action of the charge conjugation 
operator.

The computation of (\ref{Ds6}) is the aim of this paper. All energies per unit length of string
will be expressed in terms of the square of the Higgs VEV. 
\bigskip
\bigskip

\noindent{\bf 4. The Dirac Equation in an Electroweak String Background}

\bigskip

Fermions enter  the electroweak theory through the  
Lagrangian density,
\begin{equation}
 L_{fermion} = i \overline{\Psi}_{L} \gamma^{\mu} D_{\mu} \Psi_{L}
                 + i \overline{u}_{R} \gamma^{\mu} D_{\mu} u_{R}
                 + i \overline{d}_{R} \gamma^{\mu} D_{\mu} d_{R}
                 + L_{Yukawa}, \label{1.1}
\end{equation}
where,
\begin{equation}
L_{Yukawa}=- g_{u} ( \overline{\Psi}_{L} \tilde{\Phi} u_{R} 
                          +\overline{u}_{R} \tilde{\Phi}^\dagger \Psi_L )
                 - g_{d} ( \overline{\Psi}_{L} \Phi d_{R} 
                          +\overline{d}_{R} \Phi^\dagger \Psi_L ),\label{1.1.1}
\end{equation}
is the Yukawa term which is responsible for generating the masses of the 
fermions after symmetry breaking.                          

$\Phi $ is the standard Higgs doublet defined by (\ref{int4.5}) and 
$\tilde{\Phi}$ is defined to be, 
\begin{eqnarray*}
\tilde{\Phi} =
\left(
\begin{array}{c}
\phi_{0}^{*} \\ -\phi_{+}^{*}
\end{array}
\right).
\end{eqnarray*}

The left handed fermions are combined in the usual $ SU(2) $ doublet defined 
by,
\begin{eqnarray*}
\Psi_{L}=
\left(
\begin{array}{c}
u_{L} \\ d_{L}
\end{array}
\right).
\end{eqnarray*}

The covariant derivatives, which couple the fermions to the gauge fields,  
are defined by,
\begin{eqnarray*}
 D_{\mu} \Psi_{L} = ( \partial_{\mu} -i \frac{\hat g}{2} \sigma^a W^a_{\mu} 
                       -i \frac{g^{'}}{2} Y_{L} B_{\mu} ) \Psi_{L},
\quad D_{\mu} \psi_{R} = ( \partial_{\mu} - i \frac{g^{'}}{2} Y_{R} B_{\mu} ) 
                       \psi_{R},
\end{eqnarray*}                     
where $ Y_L $ and $ Y_R $ are the $U(1)_Y$ hypercharges and $\hat g $, $ g^{'} $ are the 
$ SU(2)$ and $ U(1)_Y $ coupling constants respectively. Throughout explicit 
reference is made to the up and down quarks through the notation 
$ u_{L,R} $ and $ d_{L,R} $.  Other fermions can be included by 
simply introducing the appropriate hypercharges.

Varying the Lagrangian with respect to the fermion fields 
produces four two component Dirac equations,
\begin{eqnarray*}
 i \gamma^{\mu} \partial_{\mu} u_{R}
    -q \alpha_R^u \gamma^{\mu} Z_{\mu} u_{R}
    +e e^{u} \gamma^{\mu} A_{\mu} u_{R} 
    -g_u \phi_o u_L 
        +g_u \phi_{+} d_L =0,
\end{eqnarray*}    
\begin{eqnarray*}
 i \gamma^{\mu} \partial_{\mu} u_{L}
    +q \alpha_L^u \gamma^{\mu} Z_{\mu} u_{L}
    +e e^{u} \gamma^{\mu} A_{\mu} u_{L} 
    +e^{W} \gamma^{\mu} W_{\mu}^{-} d_L
    -g_u \phi_o^* u_R 
    -g_d \phi_{+} d_R =0,
\end{eqnarray*}    
\begin{eqnarray*}    
 i \gamma^{\mu} \partial_{\mu} d_{R}
    -q \alpha_R^d \gamma^{\mu} Z_{\mu} d_{R}
    +e e^{d} \gamma^{\mu} A_{\mu} d_{R} 
    -g_d \phi_+^* u_L 
    -g_d \phi_{o}^* d_L =0,
\end{eqnarray*}    
\begin{eqnarray*}    
i \gamma^{\mu} \partial_{\mu} u_{L}
    +q \alpha_L^d \gamma^{\mu} Z_{\mu} d_{L}
    +e e^{u} \gamma^{\mu} A_{\mu} d_{L} 
    +e^{W} \gamma^{\mu} W_{\mu}^{+} u_L
    +g_u \phi_+^* u_R 
    -g_d \phi_{o} d_R =0,
\end{eqnarray*}
where,
$
 e = q \sin 2 \theta_{w} \quad{\rm and} \quad
 e^{W} = \sqrt{2} q \cos \theta_{w}.
$

Throughout we use  the Weyl representation of the Dirac matrices,
\begin{eqnarray*}
\gamma^0=
\left(
\begin{array}{cc}
0 & -1 \\
-1 & 0 
\end{array}
\right)
;
\gamma^i=
\left(
\begin{array}{cc}
0 & \sigma^i \\
-\sigma^i & 0 
\end{array}
\right)
; 
 \gamma^5=\gamma_5
\left(
\begin{array}{cc}
1 & 0 \\
0 & 1 
\end{array}
\right),
\end{eqnarray*}
with the chiral projection operators defined in the 
standard way,
\begin{eqnarray*}
 P_{R} = \frac{1}{2} ( 1 + \gamma^5 ), \quad 
 P_{L} = \frac{1}{2} ( 1 - \gamma^5 ).
\end{eqnarray*}   

Working in temporal gauge ($ W^{\pm}_o = A_o = Z_o = 0$ ) and isolating 
the time  
dependence implies that the Dirac equations read, 
\begin{equation}
i \partial_{t}
\left( \begin{array}{c}
u_R \\ u_L \\ d_R \\  d_L 
\end{array}
\right)
=
\left( \begin{array}{cccc}
-i \sigma^i D_i^{u_{R}} & -g_u \phi_o & 0 & g_u \phi_{+} \\ 
-g_u \phi_o^* &   i \sigma^i D_i^{u_{L}} & -g_d \phi_+ & e^{W} \sigma^i W_i^- \\
0 &  -g_d \phi_+^* &   -i \sigma^i D_i^{d_{R}} &    -g_d \phi_o^* \\
g_u \phi_+^* &   e^{W} \sigma^i W_i^+ &   -g_d \phi_o &  i \sigma^i D_i^{d_{L}} 
\end{array}
\right)
\left( \begin{array}{c}
u_R \\ u_L \\ d_R \\  d_L 
\end{array}
\right), \label{1.16}
\end{equation}
where $ i = 1,2,3 $ and the covariant derivatives are defined by, 
\begin{eqnarray*}
 D_{i}^{P_R} = \partial_i + i q \alpha_{R}^{p} Z_i - i e e^p A_i,
\quad
 D_{i}^{P_L} = \partial_i - i q \alpha_{L}^{p} Z_i - i  e e^p A_i.
\end{eqnarray*}
In the background of the pure Z-string the only non-zero bosonic fields are 
$ Z_i $ (where $ i=1,2$)
 and $ \phi_o $, the specific forms  of which are given in (\ref{int8}). 
The Dirac 
equation (\ref{1.16}) can be seen to  decouple into two independent 
four component equations 
for the up and down quarks.

As the Z-string solution is independent of  $z$,
the Dirac equation in the background of the Z-string can be written as,
\begin{eqnarray*}
 i \partial_t P - i C \partial_z P = 
   \hat{H}_{2+1, \lambda} P,
\end{eqnarray*}    
where $ P $ denotes a general $4$ component spinor and,
\begin{equation}
C=
\left(\begin{array}{cc}
-\sigma^3 & 0 \\
0 & \sigma^3
\end{array}
\right), \quad\label{pc}
\hat{H}_{2+1, \lambda}=
\left( \begin{array}{cc}
-i \sigma^i D_i^{P_R} & -g f(r) e^{i \lambda \theta} \\
-g f(r) e^{-i \lambda \theta} & i \sigma^i D_i^{P_L}
\end{array}
\right). \label{1.21}
\end{equation}
$\hat{H}_{2+1, \lambda}$ is the Dirac Hamiltonian defined in the plane perpendicular to the direction 
of the string.  $ i $ takes values $1, 2$, and  the  
parameter $ \lambda $ allows much of the analysis for the up and down quarks 
to be carried out using the general Hamiltonian (\ref{1.21}), with
 $ \lambda = 1$ for the up quark and $-1$ for the down quark. 
The  $ U(1)_Z $ gauge charges then satisfy,
\begin{equation}
 \alpha_L + \alpha_R = \lambda. \label{1.22}
\end{equation} 

The parameter $ \lambda $ arises as the left handed up quark couples to 
$\phi_o$ and the right handed up quark couples to the complex conjugate field 
$ \phi_o^* $, whereas the roles are reversed for the down quark with the 
left handed component coupling to $\phi_o^*$ and the right handed component 
coupling to $ \phi_o$.

The full $ 3+1 D $ Dirac Hamiltonian can be written as,
$
 \hat{H}_{3+1,\lambda} =
   \hat{H}_{2+1,\lambda} + i C \partial_z. 
$
Setting,
\begin{eqnarray*} 
P({\bf x},t)= P(x_1,x_2)e^{i E_{3+1} t} e^{-i k_z z },
\end{eqnarray*}
where $ E_{3+1} $ is the total fermionic 
energy and $ k_z $ is the momentum along the direction of the string, 
 the square of the Dirac equation becomes,
\begin{eqnarray*}
 E_{3+1,\lambda}^2 P(x_1,x_2) =
   \hat{H}_{2+1,\lambda}^2 P(x_1,x_2)
  + k_z^2 P(x_1,x_2),
\end{eqnarray*}    
where we have used $\{ C , \hat{H}_{2+1, \lambda} \} = 0$.
The problem thus reduces to solving the  $ 2+1 D $ Dirac equation, 
\begin{equation}
 E_{2+1, \lambda} P(x_1,x_2) = \hat{H}_{2+1,\lambda} P(x_1,x_2). \label{1.27}
\end{equation}

Using  $\{ C , \hat{H}_{2+1, \lambda} \} = 0$, if there is a solution of the
2+1 dimensional Dirac equation, $P(x_1,x_2) $,
with energy $ E_{2+1,\lambda} > 0 $ then $ C P(x_1,x_2) $ is easily seen to 
be a solution with energy $ - E_{2+1,\lambda}$. The particle conjugation 
operator therefore provides a $ 1 : 1 $ mapping between the positive and 
negative energy solutions of the $ 2+1 D $ Dirac equation.

Introducing polar coordinates in the plane perpendicular to the direction 
of the string, the $ 2+1 D $ Hamiltonian becomes, 
\begin{equation}
\hat{H}_{2+1, \lambda}  =\hskip-2pt
\left(\hskip-8pt
\begin{array}{cccc}
0 & i e^{-i \theta} \left({-\partial_r + \frac{i}{r} \partial_\theta \atop 
+ \alpha_R \frac{\nu (r)}{r}} \right) & -g f(r) e^{i \lambda \theta} & 0 \\
-i e^{i \theta} \left({\partial_r + \frac{i}{r} \partial_\theta\atop
+ \alpha_R \frac{\nu (r)}{r} }\right) & 0 & 0 & -g f(r) e^{i \lambda \theta} \\
-g f(r) e^{-i \lambda \theta} & 0 & 0 & i e^{-i \theta} \left({\partial_r - \frac{i}{r} 
\partial_\theta \atop+ \alpha_L \frac{\nu (r)}{r} }\right) \\
0 & -g f(r) e^{-i \lambda \theta} & -i e^{i \theta} \left({-\partial_r
- \frac{i}{r} \partial_\theta \atop + \alpha_L \frac{\nu (r)}{r}} \right) & 0
\end{array}
\hskip-8pt\right),\label{useful1}
\end{equation}
which acts on a spinor of the form,
\begin{eqnarray}
\Psi_{\lambda} =
\left( \begin{array}{c}
\psi_1^R \\ \psi_2^R \\ \psi_1^L \\ \psi_2^L
\end{array}
\right)
=\sum_{n=- \infty}^{\infty} 
\left( \begin{array}{c}
\psi_1 (r) e^{i n \theta} \\
-i \psi_2 (r) e^{i \left( n+1 \right) \theta} \\
\psi_3 (r) e^{i \left(n- \lambda \right) \theta} \\
-i \psi_4 (r) e^{i \left( n- \lambda +1 \right) \theta }
\end{array}
\right).
\label{unpmode}
\end{eqnarray}
The radial profiles of the spinor components are then determined by,
\begin{eqnarray}
(E_{3+1,\lambda} - k_z )\psi_1 & = & \left( -\partial_r - \frac{n+1}{r} +\alpha_R \frac{\nu(r)}{r}
               \right) \psi_2 - g f(r) \psi_3, \label{1.32}  \\
(E_{3+1,\lambda} + k_z )\psi_2 & = & \left( \partial_r - \frac{n}{r} +\alpha_R \frac{\nu(r)}{r}
               \right) \psi_1 - g f(r) \psi_4, \label{1.33} \\
(E_{3+1,\lambda} + k_z )\psi_3 & = & \left( \partial_r + \frac{n-\lambda+1}{r} +\alpha_L \frac{\nu(r)}{r}
               \right) \psi_4 - g f(r) \psi_1, \label{1.34} \\ 
(E_{3+1,\lambda} - k_z )\psi_4 & = & \left( -\partial_r + \frac{n-\lambda}{r} +\alpha_L \frac{\nu(r)}{r}
               \right) \psi_3 - g f(r) \psi_2. \label{a1.35}
\end{eqnarray} 

At small $r$, the Nielsen Olesen profiles give
$f(r) {\sim} r$ and $\nu (r){\sim} r^2$, which give the following
small $r$ forms for the spinors,
\begin{equation}
\Psi_1 = r^n
\left( \begin{array}{c}
1 \\ 0 \\ 0 \\ 0
\end{array}
\right) 
;
\Psi_2 = r^{-(n+1)}
\left( \begin{array}{c}
0 \\ 1 \\ 0 \\ 0
\end{array}
\right)
; 
\Psi_3 = r^{(n-\lambda)}
\left( \begin{array}{c}
0 \\ 0 \\ 1 \\ 0
\end{array}
\right)
; 
\Psi_4 = r^{-(n-\lambda+1)}
\left( \begin{array}{c}
0 \\ 0 \\ 0 \\ 1
\end{array}
\right). \label{short1}
\end{equation}

At large $r$, the Nielsen Olesen profiles give
$f(r) {\sim} 1$ and $\nu (r){\sim} 1$, which give the following
large $r$ forms for the $k_z=0$ equations of motion,
\begin{eqnarray}
E \psi_1 & = & - \left( \partial_r + \frac{\nu_{\lambda}+1}{r} \right) \psi_2 
               - g \psi_3, \label{1.41} \\
E \psi_2 & = &  \left( \partial_r - \frac{\nu_{\lambda}}{r} \right) \psi_1 
               - g \psi_4, \label{1.42} \\ 
E \psi_3 & = &  \left( \partial_r + \frac{\nu_{\lambda}+1}{r} \right) \psi_4 
               - g \psi_1, \label{1.43} \\
E \psi_4 & = &  - \left( \partial_r - \frac{\nu_{\lambda}}{r} \right) \psi_3 
               - g \psi_2, \label{1.44}
\end{eqnarray} 
where,
\begin{equation}
 \nu_\lambda = n - \lambda +\alpha_L.\label{1.45} 
\end{equation}
The spectrum of states consists of a continuum, bound states and zero modes.

If $E^2 > g^2$, the  four linearly independent regular solutions at large $r$ 
can be expressed in terms of the Bessel functions $ J_{\nu} (z) $ and 
$ N_{\nu} (z) $,
\begin{eqnarray}
\left(\begin{array}{c}
\psi_1 \\ \psi_2 \\ \psi_3 \\ \psi_4
\end{array}
\right)
& = &
a_1
\left(\begin{array}{c}
J_{\mid \nu_{\lambda} \mid} \left( k r \right) \\ 
\mp \frac{E}{k} J_{\mid \nu_{\lambda} \mid \pm 1} \left( k r \right) \\
0 \\
\pm \frac{g}{k} J_{\mid \nu_{\lambda} \mid \pm 1} \left( k r \right) \\
\end{array}
\right)
+a_2
\left(\begin{array}{c}
0 \\
\mp \frac{g}{k} J_{\mid \nu_{\lambda} \mid \pm 1} \left( k r \right) \\
J_{\mid \nu_{\lambda} \mid} \left( k r \right) \\
\pm \frac{E}{k} J_{\mid \nu_{\lambda} \mid \pm 1} \left( k r \right) 
\end{array}
\right) \nonumber \\
& & +a_3
\left(\begin{array}{c}
N_{\mid \nu_{\lambda} \mid} \left( k r \right) \\ 
\mp \frac{E}{k} N_{\mid \nu_{\lambda} \mid \pm 1} \left( k r \right) \\
0 \\
\pm \frac{g}{k} N_{\mid \nu_{\lambda} \mid \pm 1} \left( k r \right) \\
\end{array}
\right)
+a_4
\left(\begin{array}{c}
0 \\
\mp \frac{g}{k} N_{\mid \nu_{\lambda} \mid \pm 1} \left( k r \right) \\
N_{\mid \nu_{\lambda} \mid} \left( k r \right) \\
\pm \frac{E}{k} N_{\mid \nu_{\lambda} \mid \pm 1} \left( k r \right) 
\end{array}
\right), \label{1.46}
\end{eqnarray}
where, $E^2 = k^2 + g^2, \label{ms}$
is the usual mass-shell condition.
The upper signs 
correspond to the case when $ \mid \nu_{\lambda} \mid = \nu_{\lambda} $ and 
the lower signs when $ \mid \nu_{\lambda} \mid = -\nu_{\lambda} $.

If $E^2 < g^2$, the asymptotic radial solutions are modified Bessel functions
and the general solution can be written as,
\begin{eqnarray}
\left(\begin{array}{c}
\psi_1 \\ \psi_2 \\ \psi_3 \\ \psi_4
\end{array}
\right)
&=&
a_1
\left(\begin{array}{c}
I_{ \nu_{\lambda} } \left( k r \right) \\ 
- \frac{E}{k} I_{ \nu_{\lambda} + 1} \left( k r \right) \\
0 \\
 \frac{g}{k} I_{ \nu_{\lambda} + 1} \left( k r \right) \\
\end{array}
\right)
+a_2
\left(\begin{array}{c}
0 \\
- \frac{g}{k} I_{ \nu_{\lambda} + 1} \left( k r \right) \\
I_{\nu_{\lambda}} \left( k r \right) \\
\frac{E}{k} I_{\nu_{\lambda} + 1} \left( k r \right) 
\end{array}
\right) \nonumber \\
& & +a_3
\left(\begin{array}{c}
K_{\nu_{\lambda}} \left( k r \right) \\ 
\frac{E}{k} K_{ \nu_{\lambda} + 1} \left( k r \right) \\
0 \\
-\frac{g}{k} K_{\nu_{\lambda} + 1} \left( k r \right) \\
\end{array}
\right)
+a_4
\left(\begin{array}{c}
0 \\
\frac{g}{k} K_{ \nu_{\lambda} + 1} \left( k r \right) \\
K_{ \nu_{\lambda}} \left( k r \right) \\
 - \frac{E}{k} K_{\nu_{\lambda} + 1} \left( k r \right) 
\end{array}
\right), \label{1.50}
\end{eqnarray}
where, $E^2=g^2-k^2, \label{Bms}$ with $ k^2 \in (0,g^2)$. 

Only the last two independent solutions are square integrable at large $r$,
thus a physical bound state solution must match 
onto these solutions only. 

For each value of $ n $ there are two linearly independent, regular 
short distance spinors, each one  matches onto a unique 
linear combination of the four asymptotic large distance spinors. We denote the
 matching coefficients by $a_{s,l}$, where $s=1,2$ labels the short distance spinor
and $l=1-4$ labels the asymptotic spinor. To 
produce a normalisable state a suitable linear combination of the two regular 
short distance spinors must be taken such that the 
coefficients of the divergent asymptotic solutions are zero. 
Taking the linear combination, $\alpha \vec{S}_1 + \beta \vec{S}_2$,
of the two regular short distance spinors, 
square integrabiltiy implies,
\begin{eqnarray}
a_{1,1} \alpha + a_{2,1} \beta &=&  0 \nonumber \\
a_{1,2} \alpha + a_{2,2} \beta &=&  0. \label{hom1}
\end{eqnarray}
A non-trivial solution to (\ref{hom1}) exists if and only if the 
associated determinant is zero, that is,
\begin{equation}
\left|
\begin{array}{cc}
a_{1,1} & a_{2,1} \\
a_{1,2} & a_{2,2}
\end{array}
\right| \label{det1}
=0.
\end{equation}
In general, finding bound state solutions is reduced to the problem of finding the 
values of the energy which are 
roots of (\ref{det1}). 

A special case are zero modes \cite{JR}. 
Setting $E=0$ in the equations of motion,
we see that the 
system decouples into two second order systems for all $r$. 

At small $r$ the leading 
order behaviour of the two solutions to the first set is given by,
\begin{eqnarray*}
 \psi_1 \sim r^n ,\quad \psi_4 \sim r^{(n+1)}
\quad {\rm and} \quad
 \psi_1 \sim r^{-(n-\lambda)},\quad \psi_4 \sim r^{-(n-\lambda+1)}.
\end{eqnarray*} 
While the second system gives,
\begin{eqnarray*}
 \psi_2 \sim r^{-(n+1)},\quad \psi_3 \sim r^{-n},
\quad {\rm and} \quad
\psi_2 \sim r^{(n-\lambda+1)},\quad \psi_3 \sim r^{(n-\lambda)}.
\end{eqnarray*}
    
Both systems possess one regular and one irregular asymptotic solution.
Thus if both short distance solutions in one system are regular, they will match onto 
one regular and one irregular asymptotic solution. In the language used above,
$a_{1,2}=a_{2,2}=0$ (say), one constraint is automatically satisfied and there
is always a linear combination of short distance spinors that yields a square integrable
solution. Thus we have zero mode solutions.

By setting $ \lambda =1 $ it is easily seen that the only case in which there 
are two regular short distance solutions in one system is for $ n = 0 $. The
Up quark zero mode therefore has  $ n = 0 $ and $\psi_2=\psi_3=0$.
Further we see we still have a solution to the equations of motion with $\psi_2=\psi_3=0$ 
if we set $E_{3+1,\lambda}=k_z$. 
Physically this means that 
the up quark zero mode can move along the  
positive direction of the string at the speed of light. This should be 
compared with the massive bound states which are free to move in either 
direction along the string. 
This restriction on the motion of the 
zero mode has important implications for the formulation of the 
Dirac sea energy.

Similarly, by setting $ \lambda = -1 $ 
we find  the only down quark zero mode 
if $ n = -1 $. The solution has $\psi_1=\psi_4=0$  and $E_{3+1,\lambda}=-k_z$,
thus this mode is restricted
to move in the negative $z-$direction at the speed of light.

The existence of these zero mode solutions in the pure Z-string 
background is guaranteed  by an index theorem \cite{wen}. 
If the string is perturbed,  the zero modes are lifted 
and become (for small perturbations) 
low energy massive bound states \cite{nac,hong}.
 The existence of massive bound state solutions is  
 not guaranteed by any form of index theorem but depends upon 
the precise values of the parameters in 
the theory. For each set of parameters, solutions to the 
bound state condition (\ref{det1}) must  be searched for.

\bigskip
\bigskip
\noindent{\bf 5. Regulating The Fermionic Energy} 

\bigskip

We have seen that the  fermionic vacuum energy
can be calculated by   
summing over all the negative energy eigenvalues of the Dirac equation in the 
background of interest. 
Clearly this quantity is divergent: there are an infinite number 
of continuum states which make progressively larger contributions to the sum. 
 This problem is closely linked with the usual ultraviolet 
divergences which arise in quantum field theory when large values of loop 
momenta 
are integrated over. Effectively the Dirac sea energy corresponds to a 
1-loop calculation in the background of some field configuration. As with 
normal loop calculations these divergences can be consistently handled
using a renormalisation procedure. The issue of regulation is discussed
in this section and renormalisation is discussed 
in section 7.

It is useful to look in a  semi-quantitative manner at the 
nature of the ultraviolet divergences  expected. 
 The continuum energy contribution takes 
the form of an intergal over all the momentum components,  the nature of the divergences 
can be isolated by introducing a naive ultraviolet cut-off, $\Lambda$, 
\begin{eqnarray}
E_{cont}  \sim  \int^\Lambda d^3 k E \nonumber
\sim O(\Lambda^4) + O(\Lambda^2)+ O(\log \Lambda)+ \cdots \label{sq3}
\end{eqnarray}
The first term above     will cancel when two strings are compared,
leaving  quadratic ($\Lambda^2$) and 
logarithmic ($\log \Lambda$) divergences  which must be 
dealt with through the renormalisation procedure.  

The freedom to move along the length of the string also means that the 
bound states contribute to the divergences in the fermionic energy,
\begin{eqnarray}
E_{bound}  \sim  \int^\Lambda dk \sqrt{M^2_b + k^2} 
 \sim  O(\Lambda^2) + O(\log \Lambda) + \cdots \label{sq4}
\end{eqnarray}
When two string are compared it is found that there is a natural way of 
pairing the states of the two systems. The result of pairing the states  
is that the quadratic divergence in (\ref{sq4}) cancels when two 
system are compared and the resulting divergence is logarithmic. It is 
this logarithmic divergences which must be removed by the renormalisation 
procedure.

These divergences are of exactly the types expected, as  
standard Feynman diagram calculations in  the 
electroweak theory produce both quadratic and logarithmic divergences. The 
ultraviolet behaviour of a theory is associated with very large 
momentum and thus probes very small distances hence the types of the 
divergences expected should be independent of any external 
field configuration.

As discussed in section 3,  we are interested in the functional trace,
\begin{equation}
 \delta E_{fermion} = -\frac{1}{2} \mbox{Tr } \left(\sqrt {\hat{H}_1^2} 
- \sqrt {\hat{H}_2^2} \right),
\label{e1} 
\end{equation}
where the two backgrounds being compared have Dirac Hamiltonian 
operators $ \hat{H}_1 $ and $ \hat{H}_2 $. To regulate the divergences
we introduce a proper time regulator,
\begin{equation}
\delta  E_{fermion} (\tau) = \frac{1}{4 \sqrt{\pi}} \int_{\tau}^{\infty} \frac{dt}{
t^{\frac{3}{2}}} \mbox{Tr } ( e^{-t {\hat{H}_1^2}} - e^{-t {\hat{H}_2^2}} ).
\label{e2}
\end{equation}
$\tau$ acts as an ultraviolet cut off  
through the exponential suppression of high momemtum contributions and
we are interested in  the $\tau\to 0$ limit.
 It can be seen on dimensional grounds that the proper time regulator, $\tau$, is related to the
 naive ultraviolet cut off, $\Lambda$, by,
$\tau \sim \frac{1}{\Lambda^2} $.

We can see that this regulated form reproduces the original trace in the limit $\tau\to 0$ by 
using the standard integral \cite{grad},
\begin{eqnarray}
\int_\tau^\infty \frac{dt}{t^{\frac{3}{2}}} e^{- E^2 t } = 
\mid E \mid \Gamma (-\frac{1}{2}, E^2 \tau)
 \stackrel{\tau \rightarrow 0}{\rightarrow}  
\frac{2}{\sqrt{\tau}} - 2 \sqrt{\pi} \mid E \mid +O(\tau^{\frac{1}{2}}). 
\label{nn1}
\end{eqnarray}
 It is important to note that two systems must be compared so that 
the $\frac{2}{\sqrt{\tau}}$ term in (\ref{nn1}) cancels.

The spectrum of the Dirac Hamiltonian in the background of the 
Z-string consists of three distinct parts: zero modes, bound states and
continuum states,
thus the regularised fermionic energy can be written as,
\begin{eqnarray*}
 E_{fermion} (\tau) = E_{zero mode} (\tau) + E_{bound state} (\tau) + 
                        E_{cont} (\tau),
\end{eqnarray*}                                                
and each of these three parts can be handled separately.

As discussed in section 4, we split the  $3+1D$ Dirac Hamiltonian
 into a $2+1D$ Hamiltonian together with a  
contribution coming from the momentum of the fermion along length of the string,
$ {\hat{H}_{3+1}^2} = {\hat{H}_{2+1}^2} + k_{z}^2.$
 The trace must sum over all the allowed values of $k_z$, remembering that zero modes only move in one
direction along the string. We restrict the spinors to a cylindrical box of length $L$ and impose periodic boundary 
conditions in the $z-$direction, thus quantising the allowed $z$-momenta.  
Taking the limit ${L \rightarrow \infty}$,  the sum over allowed momenta becomes an integral over the 
momentum parallel to the string in the usual way.
As a  zero mode only moves in one direction along the string, its contribution 
is only half that of  a massive bound state or a continuum state.


We can now determine the zero mode contribution to the energy per unit length of the string; 
\begin{eqnarray}
 E_{zero mode} (\tau) =\sum_{zero modes} 
 \frac{1}{2 \sqrt{\pi}} \int_{\tau}^{\infty} \frac{dt}{t^{\frac{3}{2}}} 
\int_{0}^{\infty} \frac{dk}{2\pi} e^{-t k^2} 
= \frac{\alpha_{ZM}}{8 \pi} \int_{\tau}^{\infty} \frac{dt}{t^2} 
= \frac{\alpha_{ZM}}{8 \pi} \frac{1}{\tau},\label{e3}
\end{eqnarray} 
where $\alpha_{ZM}$ denotes the number of zero mode solutions to the 
$2+1D$ Dirac equation. From (\ref{e3}) it can be seen that a zero mode 
state contributes a purely manifest quadratic divergence to the 
fermionic energy.

Similarly for a massive bound state solution with  effective mass $E$, the energy contribution per unit 
length of string is given by,
\begin{eqnarray}
 E_{bound state} (\tau) &=& \frac{1}{2 \sqrt{\pi}} \int_{\tau}^{\infty} \frac{dt}{
t^{\frac{3}{2}}} \int_{-\infty}^{\infty} \frac{dk}{2 \pi} e^{-t k^2} 
e^{-t E^2}\\ \nonumber
&=& \frac{1}{4 \pi} \left( \frac{1}{\tau} 
+ E^2 \log ( E^2 \tau ) + E^2 ( C -1 ) + O(\tau) \right) , \label{e5.2}
\end{eqnarray}
where $ C $ is Eulers constant.
As with the zero modes it can be seen from (\ref{e5.2}) that each massive bound state 
contributes 
a manifest quadratic divergence to the fermionic energy. Moreover in both 
cases the manifest quadratic divergence is independent of the energy of the state. 
We shall see in section 8 that these manifest quadratic divergences 
cancel when two systems are compared as there is a natural way in which 
to pair states. 

Finally the continuum contribution to the fermionic energy per unit length 
of string reads,
\begin{equation}
 E_{cont} (\tau) = \frac{1}{4 \pi} \int_{\tau}^{\infty} \frac{dt}{t^2} 
\mbox{Tr}^+_{cont} e^{(-t \hat{H}_{2+1,\lambda}^2 )},  \label{e6}
\end{equation}
where  $\mbox{Tr}^+_{cont} $ denotes a trace over the positive 
energy part of the 
continuum spectrum of $ \hat{H}_{2+1} $ 
and we have integrated out the 
degree of freedom along the direction of the string.
  
The contributions to $E_{fermion} (\tau)$  from the massive bound states and 
zero modes are easy to compute once the fermion spectrum is known. 
The continuum 
contribution in contrast requires a significant amount of work before it is of 
a practical form for calculations. 

\bigskip
\bigskip

\noindent{\bf 6. Discretising and Recovering the Continuum}
\bigskip

In the previous section we determined an expression for the regulated contribution 
of the continuum to the total fermionic energy in the string background. 
This involves a  trace over the positive energy continuum and consequently introduces
 a sum over an infinite number of states. To carry out this sum we
 impose a boundary condition on the spinor profile functions on the surface of 
a cylinder of radius $R_o$ centred on the core of the string. This condition 
selects a discrete subset of the continuum states over which we trace.
The full continuum is recovered as $ R_o 
\rightarrow \infty $, so the  actual trace over the continuum can then be 
generated by taking this limit. This will be done 
explicitly so that the discrete sum over the discretised continuum becomes 
an integral. The use of the 
discretisation condition is merely  an intermediate step in setting up the 
sum over the continuum. This step is however critical,  because it ensures that the correct 
density of continuum states is used.

The details of the discretisation procedure are given in appendix A. 
The boundary condition we impose is  that 
the $1st$ and $3rd$ components of the spinors vanish at  $r=R_o$.  

The Dirac equation in 
the Z-string background has two regular short distant solutions which
match asymptotically onto  two linear combinations 
of the asymptotic solutions given in (\ref{1.46}). The boundary
conditions lead to a constraint of the form,
\begin{equation}
 \frac{J_{\mid \nu \mid}(k R_o)}{N_{\mid \nu \mid}(k R_o)} = -X(k),
 \label{e9}
\end{equation}  
where $X(k)$ is given in terms of the matching coefficients.

The left hand side of this constraint is approximately $\cot(kR_0)$
and consists of an  infinite number of branches of width  
proportional to $ \frac{1}{R_o} $. The density of the solutions to the 
discretisation condition thus becomes infinite as 
$ R_o \rightarrow \infty $ and so the full continuum is recovered.

Obtaining a unique solution to the discretisation condition requires careful analysis,
principally due to the occurence of singularities in $X(k)$. This issue is discussed in detail
in appendix A.

The discretisation condition can be written uniquely in the form, 
\begin{equation}
k_i R_o = k_i^o R_o + \Delta (k_i),\label{e28}
\end{equation} 
where the free momentum is defined by,
$
k_i^o R_o = i \pi,
$
 and $\Delta (k)$ is determined by matching short distance spinors to 
asymptotic spinors. The details are given in appendix A.

As, $k_i = k_i^o + O(\frac{1}{R_o})$, we have,
\begin{equation}
k_i R_o = k_i^o R_o + \Delta^\pm (k_i^o)+O(\frac{1}{R_o}). \label{int25}
\end{equation}
All of the terms on the right hand side depend only on the known 
free momentum $k_i^o$. Expanding the momentum in powers of 
$\frac{1}{R_o}$ we have,
\begin{eqnarray}
e^{-t k_i^2} = e^{-t \left[ k_i^o + \frac{1}{R_o} 
\Delta^\pm (k_i^o) 
+O(\frac{1}{R_o^2}) \right]^2}
= e^{-t (k_i^o)^2} \left[ 1 - t \frac{2 k_i^o}{R_o} \Delta^\pm (k_i^o) 
+ O(\frac{1}{R_o^2}) 
\right]. \label{int26}
\end{eqnarray}

The free momentum $k_i^o$ provides a set of evenly 
spaced nodes with spacing $\delta k = \frac{\pi}{R_o}$. In the 
large $R_0$ limit we have, 
\begin{equation}
\sum_{i=1}^{\infty} e^{ - t k_i^2 } 
= \frac{R_o}{\pi} \int_0^\infty dk e^{-t k^2} - \frac{1}{2} 
-\frac{2t}{\pi} \int_0^\infty dk k \Delta^\pm (k) e^{-t k^2} + O(\frac{1}{R_o})
. \label{int27}
\end{equation}
The first term on the right hand side of (\ref{int27}) will 
be seen to cancel when two systems are compared.

Equation \ref{int27} is exactly the 
integral representation of the continuum contribution to the 
fermionic energy we require.
We have expessed the regularised trace over the continuum states  
as an integral over the transverse momentum with the integrand defined in terms of 
matching coefficients found by solving the Dirac equation.

Finally, the total fermionic contribution to the energy of the string 
from a given angular 
momentum mode, $n$, is given by,
\begin{eqnarray}
 E^{n}_{fermion}(\tau) &=&
  \frac{\alpha_{ZM}^n}{8 \pi} \frac{1}{\tau}
+\frac{1}{4 \pi} \int_{\tau}^{\infty} \frac{dt}{t^2} \sum_{i} e^{ -t E_{n,i}^2}
\nonumber \\
& & -\frac{1}{2 \pi^2} \int_{\tau}^{\infty} \frac{dt}{t^2} e^{-t g^2}
\biggl[{\pi\over 2}- \int_{0}^{\infty} dk e^{-t k^2}\bigl[R_0-kt\sum_{\pm} \Delta^{\pm, n} (k)\bigr]
\biggl]
 \label{e30}
\end{eqnarray} 
where $ \alpha_{ZM}^n $ is the number of zero modes with mode number $ n $, 
$\{ E_{n,i} \} $ is the set of positive energy massive bound state solutions 
of the $ 2+1 D $ Dirac equation and finally $ \Delta^{\pm, n} (k) $ are the 
two functions defined by (\ref{e10}) for the pure Z-string background.

\bigskip
\bigskip
\noindent{\bf 7. Renormalisation Issues} 

\bigskip
Now we have a regulated expression for the fermionic energy shift we must
renormalise it in order to calculate the physical value of the fermionic energy.
We focus 
on the computation of the counter term in the background of the 
Z-string using the heat kernel expansion.
By calculating the counter term required to 
renomalise the contribution made by each angular momentum mode it is possible to 
renomalise each mode individually.


The highest order divergence which is expected in the 
electroweak theory is a quadratic divergence associated 
with the Higgs mass renormalisation. The remaining divergences are  
logarithmic in nature, corresponding to the wavefunction 
renormalisation of the Higgs and gauge fields together with the 
logarithmic renormalisation of the Higgs self coupling.

The physical energy of the string should be expressed in terms of 
physical parameters defined at the electroweak scale, rather than the bare parameters
that enter the Lagrangian.
Consider the sum of the classical string energy and the fermionic energy,
\begin{equation}
E_{string}=E_{classical}^{bare}+E_{fermion} (\tau), \label{rn1}
\end{equation}
where $E_{classical}^{bare}$ is the classical energy per unit length of 
the string expressed in terms of bare parameters (see (\ref{int10})) and 
$E_{fermion} (\tau)$ is the regularised fermionic energy. 
The divergences arising in the fermionic energy should be combined 
with the classical energy defined in terms of the bare parameters 
to give an expression for the classical energy defined in terms of 
the physical parameters defined at some scale $\mu$. 

In the proper time regularisation scheme 
we have,
\begin{eqnarray*}
E_{fermion} (\tau) = \frac{1}{8 \pi} \int_{\tau}^{\infty} 
\frac{dt}{t^2} \mbox{Tr} e^{-t \hat{H}^2}.
\end{eqnarray*}
The integral over the proper time $ t $ can be split into two pieces,
$\tau\to\mu$ and $\mu\to\infty$, 
where $\mu$ is some arbitrary scale (eventually $\mu$ will be set 
equal to $ M_Z^{-2}$). 
We use the heat kernel expansion 
to isolate the divergent contributions arising in the small $ \tau $ 
limit of the first integral above  and determine the counter term,
\begin{equation}
E_{CT} (\tau,\mu)=\frac{1}{8 \pi} \int_{\tau}^{\mu} 
\frac{dt}{t^2} \mbox{Tr} e^{-t \hat{H}^2} \mid_{div.}. \label{rn4}
\end{equation}

Consistency of the renormalisation scheme requires that no physical 
quantity depends on the scale $\mu$. This will be ensured by renormalisation group 
invariance, provided the  terms which comprise $E_{CT}(\tau,\mu)$ take the same 
form as those which arise in $E_{classical}^{bare}$ and 
there exist  unique expressions relating the bare parameters to 
the renormalised parameters. These conditions are satisifed in the electroweak theory 
when the proper time regulator is used \cite{Di1}.
$E_{string}$ is thus invariant under changes in 
$\mu$ and so the value of $\mu$ will be set to an energy scale 
characteristic of the electroweak theory and the parameters of the 
theory set to their values at this energy scale. Specifically, 
$
\mu=M_Z^{-2}.
$


In the electroweak theory anomolies cancel between the quark and lepton sectors.
As we consider only quarks, there is the possibility of  anomalous divergences 
appearing that can only be cancelled by including leptons.
However, as the two systems being compared are in the same topological sector,
 no anomalous divergences arise.

To compute the counter term we use the heat kernel method.
First we review the method and then we apply it in the string background.

Consider an operator, $\hat{O}$, and the associated heat equation,
\begin{equation}
\frac{\partial}{\partial t} H(t;x,y) + \hat{O}_x H(t;x,y) = 0. \label{ct2}
\end{equation}
The trace of interest is then given by \cite{ram},
\begin{equation}
\mbox{Tr } e^{-t \hat{o}} = \mbox{Tr } \int d^dx H(t;x,y). \label{ct0.3}
\end{equation}
It is important to make the distinction between the two types of traces which 
appear in (\ref{ct0.3}), the trace on the left hand side is a functional 
trace over a complete set of states whereas the trace on the right hand side 
is a standard matrix trace.

It is well known that the heat equation 
possess an asymptotic solution in 
terms of the heat kernel expansion defined by \cite{ram,ful},
\begin{equation}
H(t;x,y)=\frac{1}{(4\pi t)^\frac{d}{2}} \exp ( \frac{\mid x - y \mid^2}{4t}) 
\sum_{n=0}^{\infty} a_n (x,y) t^n. \label{ct1}
\end{equation}
(\ref{ct1}) is a power series expansion for small proper time $ t $. 
The functions $a_n (x,y)$ are the heat kernel expansion coefficients, 
their precise form will be derived later. 
 Substitution of (\ref{ct1}) into (\ref{ct0.3}) implies that, 
\begin{equation}
\mbox{Tr }e^{-t \hat{O}} = \frac{1}{(4\pi t)^\frac{d}{2}} \sum_{n=0}^{\infty} 
t^n \int d^dx \mbox{Tr } [a_n] \label{ct3.5} 
\end{equation}
where $[a_n]=a_n(x,x)$.

By direct substitution of (\ref{ct1}) into (\ref{ct2}) and equating 
powers of $t$ it is straightforward to derive a recurrence relation 
for the coefficients $a_n$. 


The counter term defined at some subtraction point, $ \mu $, involves 
a trace over the $2+1D$ Dirac Hamiltonian 
defined in the plane perpendicular to the string. 
Therefore $ d=2 $ in the heat kernel expansion and we have, 
\begin{equation}
E_{CT} (\tau ; \mu )= 
\frac{1}{32 \pi^2} \int d^2x \left(
\frac{1}{2} (\frac{1}{\tau^2} - \frac{1}{\mu^2}) Tr [a_0] + 
(\frac{1}{\tau} - \frac{1}{\mu}) Tr [a_1] 
+ \log (\frac{\mu}{\tau}) Tr [a_2] \right ), \label{ct12.3}\label{ct14}
\end{equation}
where we have kept only the divergent terms.
It is clear from (\ref{ct12.3}) that the $ [a_1] $ term corresponds to the 
quadratic counter term and the $ [a_2] $ term correspond to the 
logarithmic counter term. The quartic divergence arising from the 
$[a_0]$ term cancels when  two systems are compared. 


To derive the counter term only the 
coefficients $[a_1]$ and $[a_2]$ need to be calculated. They are found to be,
\begin{equation}
Tr \left[ a_1 \right]  =  - 4 g^2 f(r)^2,\label{ct11}
\end{equation} 
\begin{eqnarray}
Tr \left[ a_2 \right] & = & \frac{2}{3} (\alpha_L^2+\alpha_R^2) 
\left( \frac{\nu (r)^{\prime} }{r} \right) ^2 + 2 g^4 f(r)^4 
+ 2 g^2 \left( (f(r)^{\prime} )^2 + \left( \frac{f(r)}{r} \right) ^2 \right) \nonumber \\
& & + 4 g^2 \frac{\nu(r) f(r)^2}{r^2} 
+ 2 g^2 \frac{\nu(r)^2 f(r)^2}{r^2},\label{ct12}
\end{eqnarray} 
where $f(r)$ and $\nu(r)$ are the Higgs and gauge field profiles of the string respectively.

In this section the forms of both the quadratic and logarithmic counter 
terms have  been derived using the heat kernel expansion. 
The next  step is to  calculate the difference between the 
fermionic energies of two strings. In the next section such a calculation 
is carried out using a toy model which  provides  significant 
insight into the calculation of fermionic energies.

\bigskip
\bigskip
\vfil\eject
\noindent{\bf 8. The Core Model}

\bigskip


\bigskip

Before attempting a full numerical calculation of the fermionic energy 
in the Z-string background, it is very instructive to consider   
a more analytical treatment in a simplified model.  To this end we use
a core model in which simplified profiles  replace the 
actual Nielsen Olesen profiles. 
Exact expressions for the spinor wavefunctions can then be written down,
the discretisation condition can be constructed analytically and 
insight into some of its important properties can be gained. 

\bigskip

\noindent{\bf 8.1 The Model}

We use a core model defined by replacing the Nielsen-Olesen profile functions 
by,
\begin{equation}
 f(r) = \nu (r) =  \theta ( r - R_c ) \label{eq:cm1},
\end{equation}
where $ \theta (x) $ is the Heaviside step function.
Clearly these profiles do not satisfy the Nielsen-Olesen equations, however 
it is still a perfectly valid exercise to investigate fermions in 
this background. The behaviour of the core model profiles for 
$r>R_c$ is identical to the asymptotic  behaviour of the 
Nielsen-Olesen profiles, particles will therefore have their usual vacuum 
masses for $r>R_c$. While for $r<R_c$, both the Higgs and gauge fields vanish
and  all particles are massless.


The scattering of fermions off the core model string has been investigated 
in \cite{davis2} and  \cite{nag}, for completeness some of their results will be rederived.

\bigskip
\noindent{\bf 8.2 Core Model Spinors}

To find the global solutions of the Dirac equation we solve the Dirac equations
in the regions $r < R_c $ (internal) and $ r > R_c $ (external) then impose
continuity of the spinors at $ r = R_c$.
%
%
For $ r > R_c $  the general continuum ($E^2>g^2$) solution is given by
(\ref{1.46}) with, $\nu=n-\lambda+\alpha_L.$
We now drop the  subscript $\lambda$ in the definition of $\nu$  
for clarity.
Similarly, the general external bound state solution ($E^2<g^2$) 
is given by (\ref{1.50}).

The general regular internal solution reads,
\begin{equation}
\Psi_{r < R_c} =
\alpha
\left( \begin{array}{c}
J_n ( E r ) \\  - J_{n+1} ( E r) \\ 0 \\ 0 
\end{array}
\right) + \beta
\left( \begin{array}{c}
0 \\ 0 \\ J_{n-\lambda} ( E r) \\ J_{n-\lambda+1} (E r) 
\end{array}
\right). \label{cm5}
\end{equation}
For $r<R_c$ the fermions are massless,  the mass-shell condition reads $E^2=k^2$
and the argument of these solutions is $ E r $.  
These solutions are therefore suitable for use in the bound state energy region as 
well as the continuum regions.


The spinor profiles must  be continuous at $r=R_c$. For continuum states this gives  matching
conditions which allow us to determine the scattering coefficients analytically.


In the case of bound states  we only have two regular asymptotic spinors and the system of constraints is
over specified.
Non-trivial solutions to the matching conditions are then possible if and only if,
\begin{equation}
\left| \begin{array}{cc}
\begin{array}{c}
E K_{\nu+1} (k R_c) J_n (E R_c) + \\  k K_{\nu} ( k R_c ) J_{n+1} (E R_c)
\end{array} &
g K_{\nu+1} (k R_c) J_n (E R_c) \\
g K_{\nu+1} (k R_c) J_{n-\lambda} (E R_c) &
\begin{array}{c}
E K_{\nu+1} (k R_c) J_{n-\lambda} (E R_c) + \\  k K_{\nu} ( k R_c ) 
J_{n-\lambda+1} (E R_c) 
\end{array}
\end{array}
\right|
= 0 . \label{cm14}
\end{equation}
This condition  clearly puts restrictions on the energy values 
for which physical bound state wavefunctions exist and so generates a discrete
bound state spectrum.

Just as in the case of the actual Z-string there are also zero mode solutions. These occur in
the $n=0$ mode for the up quark and in the $n=-1$ mode for the down quark.
The internal solutions in the zero mode case are  simply  constants, while  the 
external solutions decay exponentially at large $ r $. 

\bigskip
\noindent{\bf 8.3 Core Model Discretisation, State Pairing and Spectral Flow}

In appendix A we discuss in  detail how to discretise the 
continuum by imposing boundary conditions on the spinor profiles at some 
large radial distance, $ r = R_o $.  The advantage of the core model is that 
analytic expressions are known for the scattering coefficients
and so an analytic expression can be written down for the 
discretisation condition.
 
Each spinor in the discretised continuum consists of a linear combination of
the two  regular internal solutions. We denote the relative amount of 
each regular internal solution in the spinor by $\gamma$.
In deriving the discretisation condition  we obtain a quadratic equation for $\gamma$ (\ref{e8}).
By direct substitution of the scattering coefficients into  
(\ref{e8.01}), (\ref{e8.02}) and (\ref{e8.1}) we find explicit forms for the coefficients and
the  quadratic reduces to, 
\begin{equation}
 \gamma^2 (g J_{n+1} J_n ) + \gamma E (J_{n+1} J_{n-\lambda} - J_{n} 
J_{n-\lambda+1}) -g J_{n-\lambda+1} J_{n-\lambda} = 0. \label{cm16}
\end{equation}
We have solutions,
\begin{equation}
\gamma=\frac{-E(J_{n+1} J_{n-\lambda} - J_n J_{n-\lambda+1}) \pm 
\sqrt{\Delta}}{2gJ_n J_{n+1}} \label{cmRoot}
\end{equation}
where the discriminant, $\Delta$, is given by,
\begin{equation}
\Delta=E^2(J_{n+1} J_{n-\lambda}- J_n J_{n-\lambda+1})^2
       +4g^2J_n J_{n+1} J_{n-\lambda+1} J_{n-\lambda}. \label{CmD}
\end{equation}
 The argument of all 
the Bessel functions in (\ref{cm16})-(\ref{CmD}) is $ E R_c $. 
There are four cases to be considered, 
corresponding to $ \mid \nu \mid = \pm\nu $ and  to the two roots of the 
quadratic equation. 
Explicitly the discretisation condition becomes,
\begin{equation} 
\frac{J_{\mid \nu \mid} ( k R_o )}{N_{\mid \nu \mid} ( k R_o) } = 
X_{\zeta\mid  \nu \mid}^{\pm} \quad {\rm for} \mid \nu \mid =  \zeta\nu 
\label{cm17} 
\end{equation}
where,
\begin{eqnarray}
X_{\zeta\mid  \nu \mid}^{\pm} & = &\hskip -0.2truecm
\frac{
E \left( J_{n+1} J_{n-\lambda} + J_{n-\lambda+1} J_n \right) 
J_{\mid \nu \mid} \mp J_{\mid \nu \mid} \sqrt{\Delta} 
- 2 \zeta k J_n J_{n-\lambda} J_{\mid \nu \mid + 1}
}
{
E \left( J_{n+1} J_{n-\lambda} + J_{n-\lambda+1} J_n \right) 
N_{\mid \nu \mid} \mp N_{\mid \nu \mid} \sqrt{\Delta} 
- 2 \zeta k J_n J_{n-\lambda} N_{\mid \nu \mid + 1}
} \label{cm18} 
\end{eqnarray}
Throughout, unless the argument is explicitly
given, the Bessel functions of integral order, associated with internal 
solutions, have argument $ E R_c $, whereas those of orders involving 
$ \mid \nu \mid $ have argument $ k R_c $.


Using the small argument 
expansions of the Bessel functions it is fairly straightforward to 
show that,

\begin{eqnarray*}
\begin{array}{ll}
 X^{\pm}_{\mid \nu \mid} (k) 
\stackrel{k \rightarrow 0}{\sim} 0 & \nu > 0 \\
X^+_{- \mid \nu \mid} (k) 
\stackrel{k \rightarrow 0}{\sim} \tan ( \mid \nu \mid \pi) & -1 < \nu  < 0 \\
X^-_{- \mid \nu \mid} (k) 
\stackrel{k \rightarrow 0}{\sim} 0 & -1 < \nu < 0 \\
X^{\pm}_{- \mid \nu \mid} (k) 
\stackrel{k \rightarrow 0}{\sim} 0 &  \nu  < -1.
\end{array}
\end{eqnarray*}


Similarly, using large argument expansions of the Bessel functions
we find as $ k \rightarrow \infty $,
\begin{eqnarray}
X^{+}_{\mid \nu \mid} & \sim & \cot \frac{\pi}{2} ( n - \mid \nu \mid ), \quad\label{l1}
X^{-}_{\mid \nu \mid} \sim \cot \frac{\pi}{2} ( n - \mid \nu \mid -\lambda), \\
X^{+}_{-\mid \nu \mid} & \sim & - \cot \frac{\pi}{2} (\mid \nu \mid - n ),\quad
X^{-}_{-\mid \nu \mid}  \sim  - \cot \frac{\pi}{2} ( \mid \nu \mid -n  -\lambda). \label{l2}
\end{eqnarray}
It is important to note that the infinite momentum behaviour of 
these functions is independent of the core radius $ R_c $. 

We can now calculate the fermionic energy difference between two core model strings with different 
core radii. This model calculation will illustrate many of the features of the full calculation, but in
a more managable context.

The eigenvalues are 
expected to be continuous functions of the core radius, $ R_c $. As a result 
a $ 1 : 1 $ correspondence should exist between the fermionic eigenvalues 
of the two strings. We first verify this by analysing a specific example.
This  provides insight into the 
physical intepretation of the singularity structure 
of the functions $ X^{\pm} (k) $.

As the core radius, $R_c$, is varied a continuous flow of the energy 
eigenvalues is expected, with the total number of fermionic states remaining constant. 
If a continuum state close to threshold loses energy and falls into the bound state region,
a new bound state solution appears.
Consequently, one of the infinitely many continuum states has become a 
bound state. It is 
essential when comparing two systems that the correct number of states are 
counted in each system so that the total number counted is the 
same in both cases, failure to do so would result in the occurrence of 
divergences which are not removed by the counter term.


For finite $ R_o $ the solutions of the  discretisation condition
are the points  where the graph of  $ \frac{J_{\mid \nu \mid} ( k R_o )}{
N_{\mid \nu \mid} ( k R_o) } $ intercepts the graphs of $ X^{\pm}_{\pm \mid 
\nu \mid} $. The function $ \frac{J_{\mid \nu \mid} ( k R_o )}{
N_{\mid \nu \mid} ( k R_o) } $ behaves 
like $ \cot ( k R_o -\frac{\mid \nu \mid \pi}{2} - \frac{\pi}{4} ) $
for $ k R_o \gg 1 $, so there are an infinite number of 
branches with a width proportional to $ \frac{1}{R_o} $. 
The singularities and zeros of
$ \frac{J_{\mid \nu \mid} ( k R_o )}{
N_{\mid \nu \mid} ( k R_o) } $ clearly occur at the zeros of 
$ N_{\mid \nu \mid} ( k R_o) $ 
and $ J_{\mid \nu \mid} ( k R_o ) $ respectively.

\begin{figure}[htp]
\begin{center}
\leavevmode
\epsfxsize=7.5cm
\epsfbox{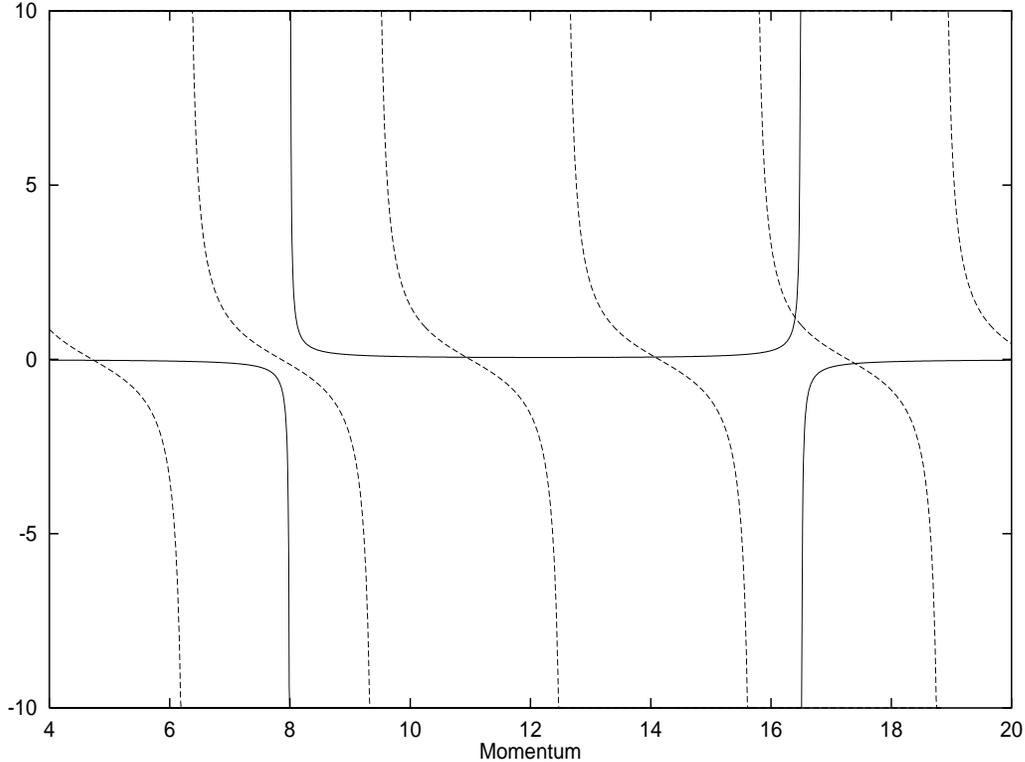}
\caption{Singularities in $X^\pm(k)$ and extra/missing states. The solid line shows
a Type 1 and a Type 2 singularity in $X^\pm(k)$ respectively. States are found at the
intesections of this line and the dashed line.} 
\label{fig:sings}
\label{fig:tp2}
\end{center}
\end{figure}   

The non-zero 
roots of $ J_{\mid \nu \mid} ( k R_o ) $ act as a suitable set with which 
the solutions of the discretisation condition (\ref{cm17}) can be matched, 
and thus provide a means of counting the continuum states.
Complications arise, as discussed in appendix A, due to the singularities
in  $ X^{\pm} (k) $. For each Type 1 singularity there is one zero of 
$ J_{\mid \nu \mid} ( k R_o ) $  for which there is no solution of the discretisation
condition [fig.\ref{fig:sings}]. Conversely, for each Type 2 singularity there is one zero of 
$ J_{\mid \nu \mid} ( k R_o ) $  for which there are two  solutions of the discretisation
condition. The number of states will be denoted by $ (1 : 1 )_{\mid \nu \mid} -q  $, 
where the notation signifies that each solution of 
the discretisation condition can be paired with a non-zero root of 
$ J_{\mid \nu \mid} ( k R_o ) $ with $q$  non-zero roots of 
$ J_{\mid \nu \mid} ( k R_o ) $ being left unpaired.

This process is independent of the specific value 
of $ R_o $. As $ R_o $ is increased the density of 
branches of $ \frac{J_{\mid \nu \mid} ( k R_o )}{N_{\mid \nu \mid} ( k R_o) }
 $ increases but the resulting state count does not change, because the 
functions $X^\pm_{\pm \mid \nu \mid} (k)$ are independent of $R_o$. 

\begin{figure}[htp]
\begin{center}
\leavevmode
\epsfxsize=7.5cm
\epsfbox{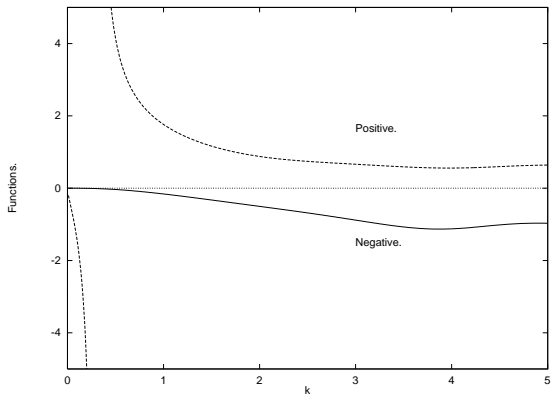}
\caption{$X^\pm(k)$ for the up quark with $n=-1$ and $R_c=1$.} 
\label{fig:sing1}
\end{center}
\end{figure}

To illustrate this  we consider two core model strings, one with $R_c=0.5$ and the other with $R_c=1$.
The parameters used are, $ \theta_w = 0.5, g=1.0$. 
Summing over the $\pm$ solutions for modes $ n=-2..2$ for both core radii, we find
$2(1:1)_{\vert\nu\vert}$ states in all cases except
for the function 
$ X^{+} (k) $ in the $ n=-1 $ mode. This difference arises because of 
the existence of a Type 1 singularity for $ R_c = 1.0 $ in $ X^{+} (k) $[fig.\ref{fig:sing1}]. 
This leaves one non-zero root of the 
appropriate Bessel function $ J_{\mid \nu \mid} ( k R_o ) $  
unpaired in the counting 
process. 

To find the total number of states we must also include any massive bound states and zero modes. 
For $ R_c = 1.0 $ there is just one massive up quark bound state and it resides in the $ n= -1 $ mode.
No massive up quark bound states exist for $ R_c = 0.5 $. 
Thus spectral flow has taken a continuum state and produced a bound state.

The inclusion of the zero modes is more subtle as they only move in one direction along the string.
A zero mode  should only 
contribute $ \frac{1}{2} $ to the state count of the mode in which it 
arises. In the $2+1D$ context the zero mode should therefore count as only 
half a state. This is an example of fermion number 
fractionization \cite{rebbi,niemi}, but it has 
been motivated by looking at the full $3+1D$ system.

As expected, the total state count is the same 
within each mode hence the total number of string states is the same. 
It is worth noting that the state counting matches within each 
mode because there is 
no spectral flow between different angular momentum sectors as the 
core radius, $ R_c, $ is varied.

The state counting procedure can be repeated using exactly the same 
techniques for the down quark. Again we find the same total state count
in both backgrounds.
 

 Having verified that the 
total number of fermionic states within a given mode does not change as 
$ R_c $ varies, we  attempt to 
compute the fermionic energy difference of the two core model strings.

\bigskip
\noindent{\bf 8.4 The Core Model Fermionic Energy Shift}

Using (\ref{e30}),  the difference in fermionic 
energy within a given mode for two systems denoted $1$ and $2$ is,
\begin{eqnarray*}
\Delta E^{n} (\tau) & = & 
\frac{(\alpha_{1,zm}^n-\alpha_{2,zm}^n)}{8 \pi} \frac{1}{\tau} 
+ \frac{1}{4 \pi} \int_{\tau}^{\infty} \frac{dt}{t^2}
\left( \sum_{i_1} e^{-t E_{n,i_1}^2} -  \sum_{i_2} e^{-t E_{n,i_2}^2} 
\right) \nonumber \\
& & -\frac{1}{2 \pi^2} \int_{\tau}^{\infty} \frac{dt}{t} e^{-t g^2} 
      \int_{0}^{\infty} dk k e^{-t k^2} \sum_{\pm} \left( 
      \Delta^{\pm, n}_1 (k) - \Delta^{\pm, n}_2 (k) \right),
\end{eqnarray*}
where the continuum contribution is encapsulated in the functions 
$ \Delta (k) $ defined by,
\begin{eqnarray*}
 \Delta(k) = \beta_{\infty} \pi + \theta_{\nu} + \cot^{-1} (X(k)) \mid_{p}^{\rm 
cont}. 
\end{eqnarray*}
The $ \theta_{\nu} $ terms cancel and the continuum contribution becomes,
\begin{eqnarray}
-  \frac{1}{4 \pi}\int_{\tau}^{\infty} \frac{dt}{t} e^{-t g^2}
\sum_{\pm}\biggl[
 (\beta_{1,\infty}^{\pm} \hskip-4pt  -\beta_{2,\infty}^{\pm}) 
 +\frac{2}{\pi} \int_{0}^{\infty} \hskip-2pt dk k e^{-t k^2} 
\left(                                      {
\cot^{-1} ( X_{1}^{\pm} (k)) \mid_{p}^{\rm cont}       \atop
-\cot^{-1} ( X_{1}^{\pm} (k)) \mid_{p}^{\rm cont}       } \right)
\biggr]. \label{cm4.39}
\end{eqnarray}

In the core model the large momentum limit of the functions $ X(k) $ is
a constant which is independent of the core radius. Thus 
the $ \cot^{-1} ( X_{1}^{\pm} (k)) \mid_{p}^{\rm cont} - 
\cot^{-1} ( X_{1}^{\pm} (k)) \mid_{p}^{\rm cont} $ term tends to zero for 
large momentum.

As discussed in appendix A, the coefficients of the first integral in (\ref{cm40}) 
can be deduced by simply counting the singularities of the functions $ X^{\pm} (k) $,
\begin{equation}
 \beta_{1, \infty}^{\pm} - \beta_{2, \infty}^{\pm} = 
\biggl(
{ \mbox{ No. Type 1 of }  X_1^\pm \atop - \mbox{ No. Type 1 of }  X_2^\pm} 
\biggr) - \biggl(
{ \mbox{ No. Type 2 of }  X_1^\pm  \atop - \mbox{ No. Type 2 of }  X_2^\pm} 
\biggr). \label{cm40} 
\end{equation}

For the up quark, the only singularity in the  functions $ X^{\pm} (k) $
occurs for the $ + $ sign case in the $ n = -1 $ mode. Thus we have, 
$ \beta_{1,\infty}^{\pm} - \beta_{2,\infty}^{\pm} =0 $, in all cases 
except for the $ + $ sign case in the $ n = -1 $ mode,
where the   type $ 1 $ singularity in $ X_1^{+, n=-1} (k) $  implies that 
$ \beta_{1,\infty}^{+} - \beta_{2,\infty}^{+} = 1 $.

In all case except for the $ + $ sign case in the $ n = -1 $ mode,
the  energy difference for each solution within a given mode reads,
\begin{eqnarray} 
\Delta E_{\pm}^{n} (\tau)  = 
-\frac{1}{2 \pi^2} \int_{\tau}^{\infty} \frac{dt}{t}  e^{-t g^2} 
\int_{0}^{\infty} dk k e^{-t k^2} \left( 
F^{\pm,n} (k)
\right), \label{cm4.41}
\end{eqnarray}
where,
\begin{equation}
F^{\pm,n}(k) = \left( 
\cot^{-1} ( X^{\pm,n}_1 (k) \mid_p ) - \cot^{-1} ( X^{\pm,n}_2 (k) \mid_p )
\right). \label{cm4.43}
\end{equation}
In the special case [$ + $ sign, $ n = -1 $] we have,
\begin{eqnarray}
\Delta E_{+}^{n=-1} (\tau)  =\frac{1}{4 \pi}\int_{\tau}^{\infty} \frac{dt}{t^2}
\biggl[ e^{-t E_{n=-1,1}^2}
-e^{-t g^2}\bigl[1-{2t\over\pi}\int_{0}^{\infty} dk k F^{+,-1} (k)\bigr]
\biggr]. 
\label{cm4.42}
\end{eqnarray}
The first term in (\ref{cm4.42}) is     the 
standard bound state contribution due to the existence of the massive 
bound state in this mode when $ R_c =1.0$. The second term is  
more subtle, it is due to the Type 1 singularity in $ X^{+,-1} (k) $ which
gives, for this particular case, 
$\beta_{1,\infty}^{+,-1} - \beta_{2,\infty}^{+,-1} = 1 $ and introduces
an extra term  from (\ref{cm4.39}). 
This term looks exactly like a massive bound state contribution except the 
state has threshold energy and the term  arises with an overall 
minus sign relative to the massive bound state. 
In the state counting the Type 1 singularity in $ X^{+,-1} (k) $ ensured that 
there are the same number of states in each mode for both string backgrounds. Spectral 
flow considerations suggest that the natural state to pair up with a massive bound state is 
a continuum state with threshold energy. Such a state would make exactly this 
contribution to the fermionic energy. These observations show  
that our formulation of the discretisation condition produces the desired state 
counting and matches states up in a consistent manner.

Whilst an individual bound state contributes a manifest quadratic 
divergence, the coefficient of this 
divergence is independent of the energy of the state. Thus this
pairing of a massive bound state with a threshold state cancels the quadratic divergences.
This is to be expected from our naive arguments using a  momentum cut-off:
while  the total fermionic energy must be quadratically divergent, the
bound states only move along the direction of the string and so they only contribute a
logarithmic divergence. 

\begin{figure}[htp]
\begin{center}
\leavevmode
\epsfxsize=7.5cm
\epsfbox{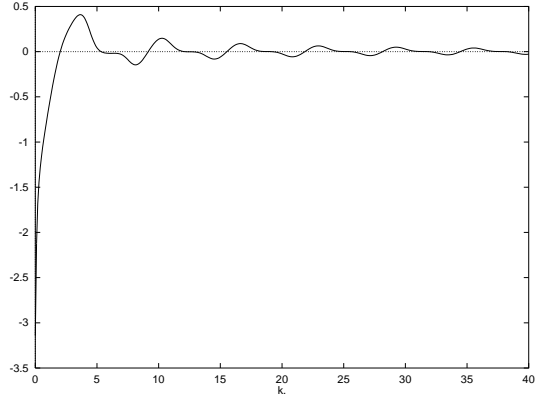}
\caption{ $ F^{-1} (k)$ for the up quark with $\theta_w = 0.5 $,
$ g=1.0 $ , $ R_{c,1}=1.0 $ and $ R_{c,2}=0.5$.} 
\label{fig:ps1}
\end{center}
\end{figure}

The total continuum contribution made by a particular mode is determined by the sum
$F^{+ , n}$ and $F^{- , n}$,
\begin{equation}
F^n(k) = \sum_{\pm} F^{\pm , n} (k). \label{cm4.45}
\end{equation}
For all modes the function $ F^n ( k )$ tends to zero at large $k$, while at small $k$
we find $ F^n ( 0 ) = 0 $ for all modes except for the $ n =-1 $ mode, where $ F^{-1} ( 0 ) = -\pi $ 
(fig.\ref{fig:ps1}). 
This is an example of Levinson's Theorem \cite{lev} 
which relates the number of bound states to the associated phase shift. In the notation used 
here, Levinson's Theorem states that  the number of massive bound states of 
the first string minus the number of massive bound states of the second 
string is equal to $-\frac{F(0)}{\pi}$.

We should now subtract the counter terms to obtain a finite energy difference.
The quadratic counter term can then be compared to the behaviour of the 
fermionic energy as $\tau$ tends towards zero and the quadratic 
renormalisation can be implemented. However there is a pathology with the logarithmic
counter term which prevents a full renormalisation of the core model. We can however
demonstrate several important features of the calculation, in particular the convergence
with mode number.

Using (\ref{ct14}), the divergent part of the quadratic 
counter term is, 
\begin{equation}
E_{Quad}(\tau) \mid_{Div} = \frac{1}{16 \pi} \frac{1}{\tau} 
\int_0^{\infty} dr r \mbox{Tr } \Delta [a_1] \label{cmct1}
=  -\frac{g^2}{4 \pi}\frac{1}{\tau}\int_0^{\infty} dr r   ( f_1(r)^2 - f_2(r)^2),\label{cmct2}
\end{equation} 
where the subscripts refer to strings $1$ and $2$. 
Using the  definition of the core model profile functions (\ref{eq:cm1}),
we have, 
 \begin{eqnarray}
E_{Quad}(\tau) \mid_{Div} 
= \frac{g^2}{8 \pi} \frac{1}{\tau} 
\left[R_{c1}^2 - R_{c2}^2\right]
= \frac{3 g^2}{32 \pi} \frac{1}{\tau}, \label{cmct4}
\label{cmct3} 
\end{eqnarray}
for the example considered ($R_{c1}=1.0$, $R_{c2}=0.5$).

The coefficient of $\frac{1}{\tau}$ in (\ref{cmct4}) should be compared with, 
\begin{equation}
\lim_{\tau \rightarrow 0 } \tau \sum_{n=-\infty}^{\infty} 
\Delta E^n (\tau), \label{cmct5}
\end{equation}
where $\Delta E^n (\tau)$ is defined by (\ref{cm4.41}) and (\ref{cm4.42}). 
The convergence of the sum over $n$ is of critical importance both fundamentally and
practically. Each $\Delta E^n (\tau)$ must be calculated numerically and hence in practice the
sum must be computed up to some finite value of $n$, $ n_{max} $,  and at some finite value of $\tau$. 
We will demonstrate that for a given value of $\tau$, providing the value of 
$n_{max}$ is sufficiently large, the higher terms in the 
sum over $n$ are indeed small and convergence is ensured. 

\begin{figure}[ht]
\centerline{
\epsfxsize=10.0cm
\epsfbox{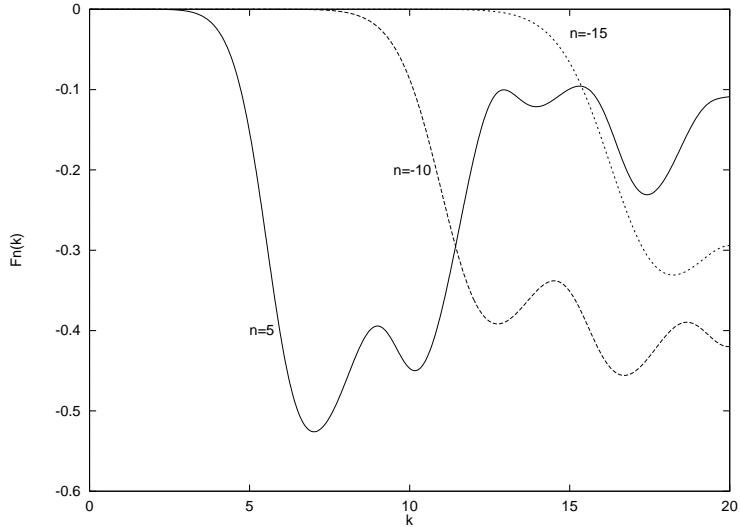}}
\caption{The functions $F^n(k)$ for the up quark with $n=5,10,15$.} \label{fig:cen1}
\end{figure}

The reason for this convergence can be seen by looking at the 
form of the energy shift functions, $F^n(k)$. Figure \ref{fig:cen1} shows the 
functions $F^n(k)$ as a function of $k$ for $n=5,10,15$ for the 
up quark, the features are however generic.
As $n$ increases, the functions 
$F^n(k)$ remain close to zero for a larger range of $k$ before the 
oscillations begin. This is due to the spinor profiles. They  are Bessel functions and
so power law suppressed until their argument is of the same magnitude as their 
order \cite{grad}. 

The proper time regulator introduces the  exponential suppression factor $ e^{-t k^2}$
into the energy integral.
As the absolute value of the mode number increases, the region where the functions 
$F^n(k)$ are  small (ie. $k < n$) becomes larger and there is increased suppression of the
overall contribution. Thus overall  we expect a suppression of the form, $e^{-\tau n^2}$.

Physically the modes with high angular momentum do not penetrate 
the core of the string and as a result are not affected 
 much by changes within the string core. 

The total fermionic contribution made by a  given angular momentum mode 
is found  by computing the momentum integrals  (\ref{cm4.41}) and (\ref{cm4.42}) 
and summing over the two cases which arise for each mode. 
The fermionic 
energy is found to decrease as the absolute value of the mode number $n$ 
increases, but more slowly as the cut-off $\tau$ becomes smaller. 
 
From a numerical point of view the momentum integral must also be 
truncated at some finite value. This value must be sufficiently large so 
that the exponential suppression has taken effect. As the value of the 
cut-off $\tau$ is taken to zero, the values of both $k$ and $n$ at which the 
exponential suppression takes effect become larger,  consequently we must extend the range of
integration in $k$ and include more modes inorder for the sum over $n$ to converge. 
For the purposes of numerical calculation, if the 
momentum integrand is required to be suppressed by a factor $tol$ from 
order unity at the truncation point, the minimum value of 
momentum at which the integral can be cut-off is given by,
\begin{equation}
k_{min} \sim \sqrt{\frac{\log tol}{-\tau}}. \label{coe4}
\end{equation}
The maximum mode number used should be of the same order as 
$k_{min}$ to ensure that the contributions made by higher $n$ modes 
are also negligible.

\begin{figure}[htp]
\begin{center}
\leavevmode
\epsfxsize=10.0cm
\epsfbox{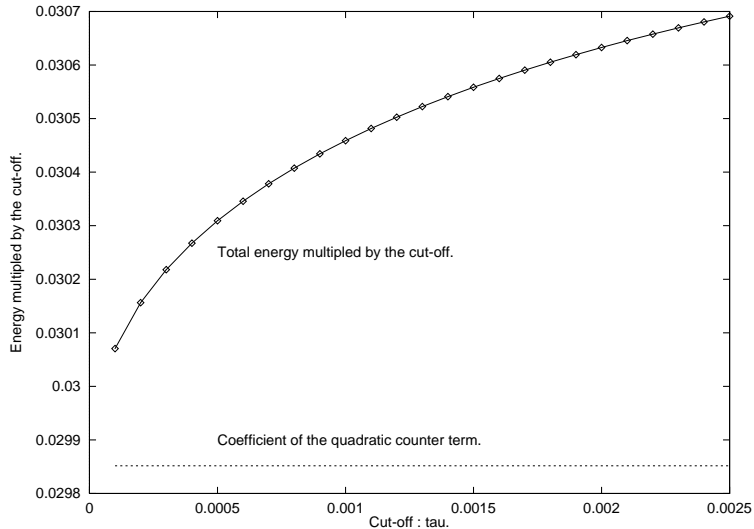}
\caption{The total regularised fermionic energy multipled by the cut-off.} 
\label{fig:ATLAST}
\end{center}
\end{figure}

It is now possible to compare the coefficients of the quadratic divergence and the quadratic counter term.
Figure \ref{fig:ATLAST} shows a plot of the total regularised fermionic 
energy multiplied by the cut-off $\tau$ as a function of $\tau$. The 400 lowest angular momentum
modes where included in the sum. The 
horizontal line is the coefficient of the quadratric counter term which 
has the value $0.02984$ for the parameters used.

It can be seen from figure \ref{fig:ATLAST} that 
$\tau E(\tau)_{fermion}$ is tending towards the coefficient 
of the quadratic counter term, shown by the horizontal line. Thus we have control
of the quadratic divergences. The curvature in 
$\tau E(\tau)_{fermion}$ becomes more significant 
as $\tau$ becomes small due to the effects of the logarithmic divergence.

Unfortunately  we cannot implement logarithmic renormalisation in the core model.
 The problem arises 
because the discontinuous nature of the core model profile functions 
implies that the energy per unit length of the string is infinite.
The Higgs field gradient term, for example, gives a contribution,
\begin{eqnarray}
\int_{0}^{\infty} dr r (f^\prime (r))^2 = 
\int_{0}^{\infty} dr r \delta^2 (r-R_c),
 \label{cmp4}
\end{eqnarray}
which is divergent.

Gradient terms of an identical form arise in the expression for the 
logarithmic counter term (\ref{ct12}), giving rise to the wavefunction 
renormalisation of both the Higgs and gauge fields. As a result the 
logarithmic counter term is not defined in the core model background and 
so logarithmic renormalisation cannot be carried out. 

The core model has provided  useful insights into the problem and tested critical aspects
of our formalism. 
It has provided insight into the process of counting the fermionic states 
in the string background,  in particular  we have demonstrated that the total number of fermionic states 
is the same for two string configurations with different profile functions.
The physical significance of the singularities in the functions $X^\pm (k)$ has been demonstrated: 
a Type 1 singularity corresponds to the occurrence of a massive bound state.
The technology of fixing the discretisation condition uniquely has been shown to ensure that the 
correct number of states are 
included in the energy sum. In particular, we have  seen that a massive bound state becomes naturally paired with a 
state at the bottom of the continuum, in accord with spectral flow considerations.
An example of Levinson's theorem has been observed, with  $ -\frac{F(0)}{\pi} $ being the difference in the number of 
massive bound states in the two systems. Finally, the  core model has demonstrated how the convergence of 
the sum over angular momentum modes arises. 
This convergence is critical for the success of the method, as numerical computations only ever allow a 
finite number of modes to be considered.

\bigskip
\bigskip
\noindent{\bf 9. The Perturbed String}

\bigskip


We now turn to the calculation of the fermionic energy difference between an actual 
Z-string and a perturbed Z-string. This will tell us what effect a  perturbation in the 
fields that cause the Z-string to be non-topological has on the fermionic energy. 
The perturbations we shall consider involve the upper
component of the Higgs field becoming non-zero. We set $ \phi_+ = g(r) $, where 
$ g(r) $ is some radial profile function which vanishes as $ r \rightarrow \infty$.

\bigskip
\noindent{\bf 9.1 The Perturbed String Dirac Equation}

The upper component of the Higgs field couples the up and down 
quarks, hence the Dirac equation will have $8$ components. 
Explicitly the $(2+1)D $ time independent Dirac equation reads, 
\begin{equation}
E
\left( \begin{array}{c}
        u_R \\ u_L \\ d_R \\ d_L
\end{array}
\right)
=
\left( \begin{array}{cccc}
      -i \sigma^i D^{u,R}_i & -g_u \phi_o & 0 & g_u \phi_+ \\
      -g_u \phi_o^* & i \sigma^i D^{u,L}_i & -g_d \phi_+ & 0 \\
      0 & -g_d \phi_+^* & -i \sigma^i D^{d,R}_i & -g_d \phi_o^* \\
      g_u \phi_+^* & 0 & -g_d \phi_o & i \sigma^i D^{d,L}_i
\end{array}
\right)
\left( \begin{array}{c}
u_R \\ u_L \\ d_R \\ d_L
\end{array}
\right),\label{pham}
\end{equation}
where all the covariant derivatives are defined in the background of 
the unperturbed 
Z-string and each of the spinors 
$ u_R, u_L, d_R, d_L $ has two components.

Using the standard Z-string ansatz and $ \phi_+ = g(r) $ together with the mode 
decomposition,
\begin{equation}
\left( \begin{array}{c}
        u_R \\ u_L \\ d_R \\ d_L
\end{array}
\right)
=
\sum_{n=-\infty}^{\infty}
\left( \begin{array}{c}
u_1^R (r) e^{i(n+1) \theta} \\
-i u_2^R (r) e^{i(n+2) \theta} \\
u_1^L (r) e^{i n \theta} \\
-i u_2^{L} (r) e^{i (n+1) \theta} \\
d_1^R (r) e^{i n \theta} \\
-i d_2^R (r) e^{i(n+1) \theta} \\
d_1^L (r) e^{i (n+1) \theta} \\
-i d_2^{L} (r) e^{i (n+2) \theta} 
\end{array}
\right)\label{umode}
\end{equation} 
we find the following set of $8$ first order, ordinary differential equations for 
the radial spinor profile functions,
\begin{equation}
E u_1^R = (-\partial_r- \frac{(n+2)}{r}+\alpha_R^u \frac{\nu(r)}{r}) u_2^R 
              -g_u f(r) u_1^L + g_u g(r) d_1^L, \label{upd1}
\end{equation}
\begin{equation}              
E u_2^R = (\partial_r- \frac{(n+1)}{r}+\alpha_R^u \frac{\nu(r)}{r}) u_1^R 
              -g_u f(r) u_2^L + g_u g(r) d_2^L,
\end{equation}              
\begin{equation}
E u_1^L = (\partial_r + \frac{(n+1)}{r}+\alpha_L^u \frac{\nu(r)}{r}) u_2^L 
              -g_u f(r) u_1^R - g_d g(r) d_1^R,
\end{equation}
\begin{equation}                            
E u_2^L = (-\partial_r + \frac{n}{r}+\alpha_L^u \frac{\nu(r)}{r}) u_1^L 
              -g_u f(r) u_2^R - g_d g(r) d_2^R,
\end{equation}
\begin{equation}              
E d_1^R = (-\partial_r - \frac{(n+1)}{r}+\alpha_R^d \frac{\nu(r)}{r}) d_2^R 
              -g_d f(r) d_1^L - g_d g(r) u_1^L,
\end{equation}
\begin{equation}
E d_2^R = (\partial_r - \frac{n}{r}+\alpha_R^d \frac{\nu(r)}{r}) d_1^R 
              -g_d f(r) d_2^L - g_d g(r) u_2^L,
\end{equation}
\begin{equation}              
E d_1^L = (\partial_r + \frac{(n+2)}{r}+\alpha_L^d \frac{\nu(r)}{r}) d_2^L 
              -g_d f(r) d_1^R + g_u g(r) u_1^R,
\end{equation}
\begin{equation}               
E d_2^L = (-\partial_r + \frac{(n+1)}{r}+\alpha_L^d \frac{\nu(r)}{r}) d_1^L 
              -g_d f(r) d_2^R + g_u g(r) u_2^R. \label{upd2}
\end{equation}              
This set of equations can easily be seen to decouple into the 
profile equations for the up and down quark in the background of the 
pure Z-string, (\ref{1.32}) to (\ref{a1.35}), if $g(r) = 0$. As $g(r)$ vanishes 
at large radial distances, the system decouples asymptotically into $2$ systems, one  for 
the up quark and one for the down quark. Thus the asymptotic solutions are given by (\ref{1.46}) 
and (\ref{1.50}) for continuum and bound states respectively.

It is important to note that the orthogonal subspace spanned by spinors with 
mode number $ n $ in (\ref{umode}) is also spanned by unperturbed  
 up quark spinors with mode number $ (n+1) $ 
and the down quark spinors with mode number $ n $ as defined by the mode 
decomposition (\ref{unpmode}).

\vfil\eject
\noindent{\bf 9.2 Perturbed String Discretisation}

Of the $8$ solutions of the Dirac profile equations (\ref{upd1}) to (\ref{upd2}), 
4 are regular at short distance. A spinor describing a physical 
 state  must consist of a linear combination of these 4 regular short 
distance spinors and will match  onto a unique linear combination of the $8$ 
regular large distance solutions. Denoting the regular short distance 
spinors by $ \vec{S}_i $ where $ i = 1, 2, 3, 4 $,  we have,
\begin{equation}
\alpha_1 \vec{S}_1 + \alpha_2 \vec{S}_2 + \alpha_3 \vec{S}_3 + \alpha_4 
\vec{S}_4
\rightarrow
\left( \begin{array}{cc}
M_u & 0 \\
0 & M_d
\end{array}
\right)
\left( \begin{array}{c}
\vec{a}_u \\
\vec{a}_d
\end{array}
\right),\label{GLC}
\end{equation}
where $ \vec{a}_{u,d}^T = ( a_1^{u,d} , a_2^{u,d} , a_3^{u,d}, a_4^{u,d} )$ 
are the projection coefficients found by matching to the large 
distance solutions. The matrices $ M_{u,d} $ are defined by the 
asymptotic Z-string continuum solutions,
\begin{equation}
M_{u,d} =
\left ( \begin{array}{cccc}
J_{\mid \nu \mid} (k R_o) & 0 & N_{\mid \nu \mid} (k R_o) & 0 \\ 
\mp \frac{E}{k} J_{\mid \nu \mid \pm 1} (k R_o) &
\mp \frac{g}{k} J_{\mid \nu \mid \pm 1} (k R_o) &
\mp \frac{E}{k} N_{\mid \nu \mid \pm 1} (k R_o) &
\mp \frac{g}{k} N_{\mid \nu \mid \pm 1} (k R_o) \\
0 & J_{\mid \nu \mid} (k R_o) & 0 & N_{\mid \nu \mid} (k R_o) \\
\pm \frac{g}{k} J_{\mid \nu \mid \pm 1} (k R_o) &
\pm \frac{E}{k} J_{\mid \nu \mid \pm 1} (k R_o) &
\pm \frac{g}{k} N_{\mid \nu \mid \pm 1} (k R_o) &
\pm \frac{E}{k} N_{\mid \nu \mid \pm 1} (k R_o) 
\end{array}
\right),\label{MM}
\end{equation}
where the appropriate subscript must be placed on both $\nu$ and $k$,
\begin{eqnarray*}
  \nu_u = n + \alpha_L^u, \quad
  \nu_d= n + 1 + \alpha_L^d, \quad
  E^2 = k_{u,d}^2 + g_{u,d}^2.
\end{eqnarray*}

By numerically solving the profile equations (\ref{upd1}) to 
(\ref{upd2}), each of the regular 
short distance spinors, $\vec{S}_i$, can evolved out to some sufficiently large distance, $R$,
and  matched to the asymptotic solutions.
If we collect the up and down quark components into 
individual 4 component spinors, $ \vec{V}_u(R) $ and $ \vec{V}_d(R) $ respectively, 
we have,
\begin{equation}
\vec{V}_u = S_u \vec{\alpha}, \label{u14}\quad
\vec{V}_d = S_d \vec{\alpha},\label{u15}
\end{equation} 
where $\alpha^T = ( \alpha_1, \alpha_2, \alpha_3, \alpha_4) $ and the scattering data is encoded in the
matrices, $S_{u,d}$,
\begin{eqnarray*}
S_{u,d}=
\left (
\begin{array}{cccc}
\psi_{1,1}^{u,d} & \psi_{1,2}^{u,d} & \psi_{1,3}^{u,d} & \psi_{1,4}^{u,d} \\
\psi_{2,1}^{u,d} & \psi_{2,2}^{u,d} & \psi_{2,3}^{u,d} & \psi_{2,4}^{u,d} \\
\psi_{3,1}^{u,d} & \psi_{3,2}^{u,d} & \psi_{3,3}^{u,d} & \psi_{3,4}^{u,d} \\
\psi_{4,1}^{u,d} & \psi_{4,2}^{u,d} & \psi_{4,3}^{u,d} & \psi_{4,4}^{u,d}
\end{array}
\right ).
\end{eqnarray*}  
Here  $\psi_{i,j}^{u,d}$ is the value of the $i$-th component of the up/down quark spinor
at $r=R$ found by evolving the $j$-th regular short distance solution.
From (\ref{GLC}) the 4 component spinors  $ \vec{V}_u $ and $ \vec{V}_d $ 
can be written as,
\begin{equation}
\left( \begin{array}{c}
\vec{V}_u \\ \vec{V}_d
\end{array}
\right) = 
\left( \begin{array}{cc}
M_u & 0 \\
0 & M_d
\end{array}
\right)
\left( \begin{array}{c}
\vec{a}_u \\
\vec{a}_d
\end{array}
\right). \label{NGLC}
\end{equation}
By eliminating $\vec{\alpha}$ using (\ref{u14}), we can express $  \vec{a}_u$ or $\vec{a}_d$ in terms of 
the other,  
\begin{equation}
\vec{a}_d = M \vec{a}_u, \label{u17}
\end{equation} 
where,
\begin{equation}
M = M_d^{-1} S_d S_u^{-1} M_u.\label{u18}
\end{equation} 
Thus only $4$ out of the $8$ asymptotic projection 
coefficients are linearly independent. 
We arbitrarily treat the components of  $  \vec{a}_u$ as our arbitrary coefficients.
These degrees of freedom are of course related to the components of $\vec{\alpha}$. 
In fact, there are only $3$ degrees of freedom  because the spinor is only defined 
up to an overall normalisation constant.

As with the pure Z-string, the next step is to 
discretise the continuum by imposing  boundary conditions on the spinor profiles
at $r=R_o$. We {\it double up} the conditions used for the pure Z-string,
 this allows much of the technology 
developed in section 6 to be applied to the  perturbed string. Specifically   
we set the first, third, fifth and seventh components of the $8$ component 
spinor to zero at $r=R_o$. Clearly as the perturbation is switched off 
these conditions become  identical to those used in the 
case of the unperturbed string.     

Applying the boundary conditions at $R_0$ we find,
\begin{equation}
\frac{J_{\mid \nu_u \mid}(k_u R_o)}{N_{\mid \nu_u \mid}(k_u R_o)}=
-\frac{a^u_3}{a^u_1}=-\frac{a^u_4}{a^u_2}, \label{u23}
\end{equation} 
and, 
\begin{equation}
\frac{J_{\mid \nu_d \mid}(k_d R_o)}{N_{\mid \nu_d \mid}(k_d R_o)}=
-\frac{a^d_3}{a^d_1}=-\frac{a^d_4}{a^d_2}.\label{u24}
\end{equation} 
Individually (\ref{u23}) and (\ref{u24}) are identical in nature 
to the discretisation 
conditions which arose in the case of the pure Z-string. 
However (\ref{u23}) and (\ref{u24}) must be solved simultaneously due to  the non-trivial 
coupling between the up and down quarks induced by the upper component of 
the Higgs field.

We seek a discretisation condition of the form, 
\begin{eqnarray*}
\frac{J_{\mid \nu \mid}(k_u R_o)}{N_{\mid \nu \mid}(k_u R_o)}=
-X(k),
\end{eqnarray*} 
in order to employ  the methology  developed in section 6. 
From (\ref{u17}) we have,
\begin{equation}
\frac{a_3^d}{a_1^d}=
\frac{M_{3,1} \frac{a_1^u}{a_2^u}+M_{3,2}+M_{3,3}\frac{a_1^u}{a_2^u}
\frac{a_4^u}{a_2^u}+M_{3,4}\frac{a_4^u}{a_2^u}}
{M_{1,1} \frac{a_1^u}{a_2^u}+M_{1,2}+M_{1,3}\frac{a_1^u}{a_2^u}
\frac{a_4^u}{a_2^u}+M_{1,4}\frac{a_4^u}{a_2^u}},\label{u25}
\end{equation} 
and
\begin{equation}
\frac{a_4^d}{a_2^d}=
\frac{M_{4,1} \frac{a_1^u}{a_2^u}+M_{4,2}+M_{4,3}\frac{a_1^u}{a_2^u}
\frac{a_4^u}{a_2^u}+M_{4,4}\frac{a_4^u}{a_2^u}}
{M_{2,1} \frac{a_1^u}{a_2^u}+M_{2,2}+M_{2,3}\frac{a_1^u}{a_2^u}
\frac{a_4^u}{a_2^u}+M_{2,4}\frac{a_4^u}{a_2^u}},\label{u26}
\end{equation} 
where $a_3^u$ has also been eliminated using (\ref{u23}). 

Implementing part of (\ref{u24}) by equating (\ref{u25}) with (\ref{u26}), we find
a quadratic equation for $ \frac{a_1^u}{a_2^u} $,    
thus we can consider $ \frac{a_1^u}{a_2^u}$ to be  a function of $\frac{a_4^u}{a_2^u}$.
Substituting the solutions of this quadratic  into (\ref{u25}) and (\ref{u26}), the 
 energy of the continuum states and the ratio $\frac{a_4^u}{a_2^u}$ are the 
only remaining independent degrees of freedom. We must now impose the 
2 remaining constraints,
\begin{equation}
\frac{J_{\mid \nu_u \mid}(k_u R_o)}{N_{\mid \nu_u \mid}(k_u R_o)}=
-\frac{a^u_4}{a^u_2} \label{u33}
\quad {\rm and}\quad
\frac{J_{\mid \nu_d \mid}(k_d R_o)}{N_{\mid \nu_d \mid}(k_d R_o)}=
-\frac{a^d_3}{a^d_1},\label{u34}
\end{equation}
to determine the allowed energy $E$ and the ratio  $\frac{a_4^u}{a_2^u}$.

 A solution to this system has only been found for degenerate masses, 
$g_u=g_d$ and hence $k_u=k_d$. For degenerate masses and $ k R_o \gg 1 $, 
using the standard large distance expansions of the Bessel functions, we have,

\begin{eqnarray*}
\cot(kR_o - \theta_u) = -\frac{a_4^u}{a_2^u},\quad
\cot(kR_o - \theta_d) = -\frac{a_3^d}{a_1^d},
\end{eqnarray*}
where,
\begin{equation}
\theta_{u,d} = \frac{\pi}{2} ( \mid \nu_{u,d} \mid + \frac{1}{2} ).
\label{dd}
\end{equation}
Using standard trigonometric identities to eliminate $kR_o$ gives,
\begin{equation}
\frac{\frac{a_4^u}{a_2^u}+\tan \theta_u}{\frac{a_4^u}{a_2^u} \tan \theta_u -1}
-\frac{\frac{a_3^d}{a_1^d}+\tan \theta_d}{\frac{a_3^d}{a_1^d} \tan \theta_d -1}
   =0.\label{u38}
\end{equation}
The only independent variable in (\ref{u38}) is the 
ratio  $\frac{a_4^u}{a_2^u}$ and we have
a one dimensional root finding problem. This process, although
conceptually straightforward, requires a significant amount of  
computational effort due to the singularities found in (\ref{u38}). 
Numerically solving (\ref{u38}) for a range of continuum 
energy values produces functions $\frac{a_4^u}{a_2^u} (E)$ which 
can then be used in (\ref{u23}) to provide a discretisation 
condition of a similar nature to that of the pure Z-string.


For a given value of the energy, 4 solutions 
of (\ref{u38}) are expected. Denoting 
these  solutions  by $ X^\alpha (k) $ where $ \alpha =1,2,3,4 $,  for 
$ k R_o \gg 1 $ the discretisation condition reads,
\begin{equation}
\cot ( k R_o - \theta_u ) = -X^\alpha (k), \label{u43}
\end{equation} 
which can be inverted using (\ref{e28}) to give, 
\begin{equation}
k_i R_o = i \pi + \beta_\infty^\alpha \pi + \theta_u + \cot^{-1} (- X^\alpha (k)) 
\mid_P^{cont} ,\label{u40}
\end{equation}   
with the value of $ \beta_\infty^\alpha $ given by (\ref{beg}).

Using the analysis of section 6, the continuum contribution to the fermionic energy of the
perturbed string made by a given angular momentum mode is,
\begin{eqnarray}
E^n_{cont} (\tau) & = & 
\frac{R_o}{\pi^2} \int_{\tau}^{\infty} \frac{dt}{t^2} e^{-t g^2} 
\int_{o}^{\infty} dk e^{-t k^2} 
-\frac{1}{2 \pi} \int_{\tau}^{\infty} \frac{dt}{t^2} e^{-t g^2} \nonumber \\
&-& \frac{1}{4 \pi} \sum_{\alpha=1}^4 \beta_{\infty}^{n,\alpha} 
\int_{\tau}^{\infty} \frac{dt}{t^2} e^{-t g^2} 
-\frac{\theta_u}{\pi^2} \int_{\tau}^{\infty} \frac{dt}{t^2} e^{-t g^2} \nonumber \\
&-& \frac{1}{2 \pi^2} \int_{\tau}^{\infty} \frac{dt}{t} e^{-t g^2} 
\int_{0}^{\infty} dk k e^{-t k^2} \sum_{\alpha=1}^4 
\cot^{-1} (-X^{n,\alpha} (k)) \mid_p^{cont}.\label{u41}
\end{eqnarray} 
As in the case of the unperturbed string, the first two terms
cancel when two systems are compared.

We must also consider the bound states of the perturbed string by looking for
 suitable linear combinations of the regular short distance spinors  that match, 
at large radial distances, to spinor profiles that decay exponentially. In the  
perturbed string there are 4 regular short distance spinors in each mode, we denote these by  $\vec{S}_i$. 
The perturbations we are considering  are localised to the string core so, just as in the case of the 
continuum, the 8 linearly independent 
large distance solutions are the same as those which arose in the 
case of the unperturbed string (\ref{1.50}).

As in the continuum, we have a doubling up of states and matching conditions.
The existence of a massive bound state solution requires that 
the coefficients of the exponentially divergent solutions vanish 
when a suitable short distance spinor is evolved to large 
distance. Taking a general linear 
combination of the 4 regular short distance spinors gives the 
bound state condition,
\begin{equation} 
\alpha_i a_{i,1}^u = \alpha_i a_{i,2}^u = \alpha_i a_{i,1}^d = 
\alpha_i a_{i,2}^d = 0,\label{pb3}
\end{equation} 
where the summation convention is used and the asymptotic states are defined in (\ref{1.50}). 
These  conditions  lead to the determinant equation,
\begin{equation}
\left| \begin{array}{cccc}
a_{1,1}^u & a_{2,1}^u & a_{3,1}^u & a_{4,1}^u \\
a_{1,2}^u & a_{2,2}^u & a_{3,2}^u & a_{4,2}^u \\
a_{1,1}^d & a_{2,1}^d & a_{3,1}^d & a_{4,1}^d \\
a_{1,2}^d & a_{2,d}^d & a_{3,2}^d & a_{4,2}^d 
\end{array}
\right|
= 0 .\label{pb4}
\end{equation} 
We find two bound states in the perturbed string background,
\begin{eqnarray}
E^{n=-1}&=&0.0505,\quad
E^{n=0}=0.9992.
\end{eqnarray}
 $ n=-1 $ corresponds to the mode in which the up and 
down quark zero modes reside if the perturbation is switched off,
thus, as  first shown in \cite{nac},  the 
perturbation lifts the energy of the zero modes to form
a low energy bound state. The index theorem \cite{wen} which ensures 
that there 
are zero mode solutions in the background of the pure Z-string no longer 
applies because the perturbation in the upper component of the Higgs field 
is exactly one of the fields which makes the electroweak string 
non-topological. In terms of the spectral flow of fermion energy levels, 
as the perturbation in the upper component of the Higgs field is turned on, the 
up and down quark zero modes, which are restricted to move in opposite 
directions along the length of the string, are combined to from a low 
energy massive bound state. The number of fermionic degrees of freedom associated 
with the zero modes is therefore unchanged. 
Similar behaviour has been observed in the context of GUT string zero modes mixing with
massless modes \cite{d2perk}.

Denoting the set of massive bound state solutions in a given mode 
 by $ \{ E_{n,i} \} $,  the total contribution made by the mode 
to the fermionic energy becomes,  
\begin{eqnarray}
E^{n}_{fermion} (\tau) && =
\frac{R_o}{\pi^2} \int_{\tau}^{\infty} \frac{dt}{t} e^{-t g^2} 
\int_{0}^{\infty} dk e^{-t k^2} 
-\frac{1}{2 \pi} \int_{\tau}^{\infty} \frac{dt}{t^2} e^{-t g^2} \nonumber \\
& & + \frac{1}{4 \pi} \int_{\tau}^{\infty} \frac{dt}{t^2} 
\sum_{i} e^{-t E_{n,i}^2} 
-\frac{1}{4 \pi} \sum_{\alpha=1}^4 \beta_{\infty}^{n,\alpha} 
\int_{\tau}^{\infty} \frac{dt}{t^2} e^{-t g^2} 
-\frac{\theta_u}{\pi^2} \int_{\tau}^{\infty} \frac{dt}{t^2} e^{-t g^2} 
\nonumber \\
& & - \frac{1}{2 \pi^2} \int_{\tau}^{\infty} \frac{dt}{t} e^{-t g^2} 
\int_{0}^{\infty} dk k e^{-t k^2} \sum_{\alpha =1 }^4 
\cot^{-1} (-X^{n,\alpha} (k)) \mid_{p}^{cont}. \label{pe1}
\end{eqnarray}

\bigskip
\noindent{\bf 9.3 The Perturbed String Counter Term}

The calculation of the counter term for the perturbed string 
proceeds along the same lines as the corresponding calculation for the 
pure Z-string. The 4 component Z-string
Dirac Hamiltonian is replaced by the 8 component Dirac Hamiltonian 
defined in (\ref{pham}). The counter term is 
the sum of the individual up and down quark counter terms in the 
background of the pure 
Z-string (as defined by (\ref{ct11}), (\ref{ct12}) and (\ref{ct14})) together 
with some modification arising from the perturbation.
 
By squaring the Hamiltonian in (\ref{pham}) and writing it in the form,
\begin{equation}
{\hat H}^2=-(\partial_i+\Gamma_i)(\partial_i+\Gamma_i) +a(x),
\end{equation}
we can apply the formalism for the heat kernel expansion developed in section 7.
Explicitly we have, 
\begin{equation}
\Gamma_i = iq Z_i{\hskip 2pt} \mbox{diag} \left( \alpha_R^u I_{2 \times 2} , 
-\alpha_L^u I_{2 \times 2} ,\alpha_R^d I_{2 \times 2} ,
-\alpha_L^d I_{2 \times 2} \right), \label{pgam}
\end{equation}
and,
\begin{equation}
a(x)=\left(
\begin{array}{cc} 
a_u(x) & a_I(x) \\
a_I^\dagger (x) & a_d(x) 
\end{array}\right), \label{pa}
\end{equation}
where,
\begin{equation}
a_u(x)=\left(
\begin{array}{cc}
-iq \alpha_R^u (\partial_i Z_i - \sigma^j \sigma^i \partial_i Z_j) &   i g_u \sigma^i \partial_i (f(r) e^{i \theta})    \\
 + g_u^2 f(r)^2 + g_u^2 g(r)^2                                     &    -g_u q \sigma^i Z_i f(r) e^{i \theta}          \\
                                                                   &                                                      \\
-ig_u \sigma^i \partial_i (f(r) e^{-i \theta})           & iq \alpha_L^u (\partial_i Z_i - \sigma^j \sigma^i \partial_i Z_j)\\
 -g_u q \sigma^i Z_i f(r) e^{-i \theta}                  & + g_u^2 f(r)^2 + g_d^2 g(r)^2 \\
\end{array}
\right), \label{au}
\end{equation}
\vskip 3mm
\begin{equation}
a_d(x)=\left(
\begin{array}{cc}
-iq \alpha_R^d (\partial_i Z_i - \sigma^j \sigma^i \partial_i Z_j) & i g_d \sigma^i \partial_i (f(r) e^{-i \theta})  \\
+ g_d^2 f(r)^2 + g_u^2 g(r)^2                                      & +g_d q \sigma^i Z_i f(r) e^{-i \theta}          \\
                                                                   &                                                 \\
-ig_d \sigma^i \partial_i (f(r) e^{i \theta})    &  iq \alpha_L^d (\partial_i Z_i - \sigma^j \sigma^i \partial_i Z_j)\\
+g_d q \sigma^i Z_i f(r) e^{i \theta}            &   + g_d^2 f(r)^2 + g_d^2 g(r)^2                                  \\  
\end{array}
\right), \label{ad}
\end{equation}
and,
\begin{equation}
a_I(x)=\left( 
\begin{array}{cc}
0                                                  & g_u q g(r) \sigma^i Z_i (\alpha_R^u+\alpha_L^d)    \\
                                                   &               - ig_u \sigma^i\partial_i g(r)      \\
                                                   &                                                    \\
-ig_d g(r) \sigma^i \partial_i g(r)                &                                                       \\
- q g_d g(r) \sigma^i Z_i (\alpha_R^d +\alpha_L^u) & (g_d^2-g^2_u) g(r) f(r) e^{-i \theta}            \\
\end{array}
\right). \label{aI}
\end{equation}
After a similar,  but more lengthy, calculation to that presented in appendix B, 
we find  the heat kernel coefficients in the 
perturbed string background are,
\begin{equation}
\mbox{Tr} [a_1] = -4 (g_u^2+g_d^2) ( f(r)^2 + g(r)^2), 
\label{pco1}
\end{equation}
and,
\begin{eqnarray}
\mbox{Tr} [a_2] = & & 
\frac{2}{3} (\alpha_R^{u^2}+\alpha_L^{u^2}+\alpha_R^{d^2}+\alpha_L^{d^2}) 
\left( \frac{\nu(r)^\prime}{r} \right)^2 \nonumber \\
&+& (g_u^2 f(r)^2+ g_d^2 g(r)^2)^2 + (g_u^2 f(r)^2 + g_u^2 g(r)^2)^2 \nonumber \\
&+&(g_d^2 f(r)^2 + g_u^2 g(r)^2)^2 + (g_d^2 f(r)^2 + g_d^2 g(r)^2)^2
\nonumber \\
&+& 2 (g_u^2+g_d^2) \left( (f(r)^\prime)^2 + \left(\frac{f(r)[\nu(r)+1]}{r}\right)^2 \right) 
+ 2 (g_d^2 - g_u^2)^2 g(r)^2 f(r)^2 \nonumber \\
&+&2 g_u^2 \left( (g(r)^\prime)^2+(\alpha_R^u+\alpha_L^d)^2 
\left(\frac{\nu(r) g(r)}{r} \right)^2 \right) \nonumber \\ 
&+&
2 g_d^2 \left( (g(r)^\prime)^2+(\alpha_R^d+\alpha_L^u)^2 
\left(\frac{\nu(r) g(r)}{r} \right)^2 \right).\label{pco2}
\end{eqnarray}
We see that 
(\ref{pco1}) and (\ref{pco2}) reduce to the sum of the individual contributions 
made by the up and down quark in the background of the unperturbed string when 
the perturbation is set to zero.
 
Having found expressions for both the fermionic energy of the perturbed string 
and the corresponding counter term, these results can be combined with 
the corresponding calculations for the unperturbed string to compare 
the fermionic energy of a perturbed string to that of an unperturbed string.

\bigskip
\noindent{\bf 9.4 The Change in the Ground State Energy}

In order to obtain a well behaved mode decomposition of the energy difference, we must
examine the angular dependence of the spinors.
The space spanned by the $ n$th modes of the perturbed string energy eigenstates (\ref{umode}) 
is also spanned by the set formed by the union of the $ (n+1)$th  up quark modes and the 
$ n$th down quark modes in the background of the 
unperturbed string, as defined by (\ref{unpmode}). Physically this means 
that the spectral flow caused by the perturbation  only mixes the $ (n+1)$th
up quark mode and the $ n$th down quark mode. Thus the energy 
difference is found by subtracting the sum of the contributions made by the $ (n+1)$th 
up quark mode and the $ n$th down quark mode in the background of the 
unperturbed string from the contribution made by the $ n$th perturbed 
string mode. Matching modes in this manner ensures that the 
fermionic  states of the two systems can be put into a $1:1$ correspondence.

This coupling of modes is consistent with the mixing of the up quark zero mode ($n=-1$)
and the down quark zero mode ($n=0$).

Given this coupling of the modes, it is convenient to make the following mapping on the unperturbed up quark 
mode number which was used in section 4,
\begin{eqnarray*}
n_u \Rightarrow n -1 .
\end{eqnarray*}
Again we split the energy difference  
into a bound state contribution and a continuum contribution. 
For the bound states we have,
\begin{eqnarray}
\delta E_{bound}^n (\tau)&=&- \frac{(\alpha_{ZM,d}^n+\alpha_{ZM,u}^n)}{8 \pi \tau}
+\frac{1}{4 \pi} 
\sum_{i}\biggl( (E_{n,i}^P)^2 \Gamma (-1, \tau (E_{n,i}^P)^2) \hfill\nonumber \\
&&-(E_{n,i}^d)^2 \Gamma (-1, \tau (E_{n,i}^d)^2)           
-(E_{n,i}^u)^2 \Gamma (-1, \tau (E_{n,i}^u)^2) \biggr) \nonumber \\
&&-\frac{1}{4 \pi} \left( 
\sum_{\alpha=1}^4 \beta_{\infty,P}^{n,\alpha} - \sum_{\pm} \beta_{\infty,d}^{n,\pm} 
- \sum_{\pm} \beta_{\infty,u}^{n,\pm} \right) g^2 \Gamma(-1,g^2 \tau),
\label{pb}
\end{eqnarray}
where  $\{ E_{n,i}^P \}$, $\{ E_{n,i}^d \}$ and $\{ E_{n,i}^u \}$ are the set of
positive energy massive bound state solutions for the perturbed string, down 
quark and up quark respectively with mode number $ n $ . 
$\alpha_{ZM,u}^n $ and $\alpha_{ZM,d}^n $ denote the number of up and 
down quark zero mode solutions respectively in the unperturbed background with mode number $ n $. 
The values of the integers $ \beta_{\infty} $ required to fix the discretisation condition uniquely at large momentum, 
are determined by counting the singularities in the functions $ X (k) $ and using (\ref{beg}).

The continuum contribution to the energy difference is given by,
\begin{eqnarray}
\delta E_{cont}^n (\tau)  & = & 
-\frac{1}{2 \pi^2} (\theta_{\nu_u} - \theta_{\nu_d}) 
\int_{\tau}^{\infty} \frac{dt}{t^2} e^{-t g^2} \nonumber \\
& & -\frac{1}{2 \pi^2} \int_{\tau}^{\infty} \frac{dt}{t} e^{-t g^2} 
\int_{0}^{\infty} dk k e^{-t k^2} \left(
\sum_{\alpha=1}^4 \cot^{-1} (-X^{n,\alpha}(k)) \mid_p^{cont} \right. \nonumber \\
& & \left. -\sum_{\pm} \cot^{-1} (-X^{n,d}(k)) \mid_p^{cont}
-\sum_{\pm}^4 \cot^{-1} (-X^{n,u}(k)) \mid_p^{cont}
\right).\label{com1}
\end{eqnarray}
The second term in (\ref{com1}) is of a familiar form,  while the first 
is a new feature of the perturbed string. 
Treating $ \frac{a_4^u}{a_2^u} $ as the independent variable gives
 a discretisation condition of the form in (\ref{u40}).
This has  the same form as the discretisation condition for the  
 up quark in the unperturbed string background. Had we used $  \frac{a_3^d}{a_1^d} $ 
as the independent variable instead,  the discretisation condition would have taken the form,
\begin{eqnarray*}
k_{i} R_o = i \pi +\beta_{\infty} \pi + \theta_{\nu_d} 
+\cot^{-1} (- \stackrel{~}{X(k)} ) \mid_p^{cont} , 
\end{eqnarray*}
which is of the same form as the discretisation condition of the down quark 
in the unperturbed string background. Thus, by working with $ \frac{a_4^u}{a_2^u} $,
we are taking the up quark states of the unperturbed string as our reference points in 
both sectors. The first term in (\ref{com1}) represents the shifts of the 
down quark states of the unperturbed string relative to these up quark reference points.
There are of course corresponding factors within the perturbed string phase shifts, so that
overall the energy difference is independent of the reference points used.
It is most natural to absorb this term into the integral over the continuum momentum. 
Using (\ref{dd}) it is easily seen that, 
\begin{equation}
\theta_{\nu_u} - \theta_{\nu_d} = \frac{\pi}{2} ( \mid \nu_u \mid - 
\mid \nu_d \mid ), 
\end{equation}
and so,
\begin{eqnarray}
\delta E^n_{cont} = 
  -\frac{1}{2 \pi^2} \int_{0}^{\infty} dk k \Gamma (0, \tau(k^2+g^2)) F^n(k),
\label{com2}
\end{eqnarray}
where,
\begin{eqnarray}
F^n(k) & = & \sum_{\alpha=1}^2 \cot^{-1} (-X^{\alpha}_{p,n}(k)) \mid_p^{cont} 
-\sum_{\pm} \cot^{-1} (-X^{\pm}_{d,n}(k)) \mid_p^{cont} \nonumber \\
& & -\sum_{\pm} \cot^{-1} (-X^{\pm}_{u,n}(k)) \mid_p^{cont}
-\pi ( \mid \nu_d \mid - \mid \nu_u \mid ). \label{pF}
\end{eqnarray}

After some simplification, the counter term for the energy difference is given by,
\begin{equation}
E_{CT} (\tau, \mu) = \frac{1}{16 \pi} 
\int_{0}^{\infty} dr r \left( 
(\frac{1}{\tau}-\frac{1}{\mu}) \Delta [a_1] 
+ \log (\frac{\mu}{\tau})  \Delta [a_2] \right),\label{coct1}
\end{equation}
where,
\begin{equation}
\Delta [a_1] = -4 (g_u^2+g_d^2) g(r)^2, \label{coct2}
\end{equation}
and,
\begin{eqnarray}
\Delta [a_2] & = & 2(g_u^4+d_d^4) (g(r)^4 +2f(r)^2 g(r)^2) 
+2(g_u^2+g_d^2) (g^\prime (r))^2 \nonumber \\
& & +2\left( g_u^2(\alpha_R^u+\alpha_L^d)^2+g_d^2(\alpha_R^d+\alpha_L^u)^2
\right) \left( \frac{\nu(r) g(r)}{r} \right)^2.\label{coct3}
\end{eqnarray}

The degenerate mass case under consideration is obtained by 
simply putting $ g_u = g_d = g $ in (\ref{coct2}) and (\ref{coct3}). 

\begin{figure}[htp]
\begin{center}
\leavevmode
\epsfxsize=7.5cm
\epsfbox{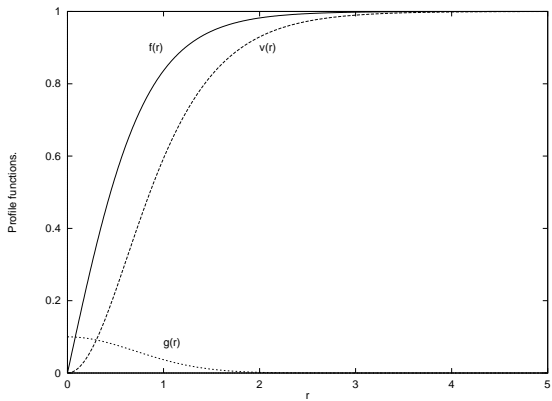}
\caption{The background profile functions used.}
\label{fig:bkg}
\end{center} 
\end{figure}

The counter term in (\ref{coct1}) still contains the arbitrary scale 
$\mu $. As discussed in section 7, the 
total energy is independent of the particular value of $\mu $  
because the parameters in the theory also depend upon $\mu $ in exactly 
the manner to ensure that physical quantities are independent of 
$\mu $. The value of $\mu $ is set by the 
typical energy scale in the problem, which we take to be the 
mass of the Z boson. By fixing the vacuum expectation value of the Higgs field 
 to be $\nu = 247 \mbox{ GeV}$, we have,
\begin{equation}
M_z = \frac{76.9 \mbox{ GeV}}{\mid \sin 2 \theta_w \mid}. \label{fin1}
\end{equation}
Taking $\theta_w = 0.5$, we have $ M_z \approx 0.311$ in units of the vacuum expectation 
value, giving,
\begin{equation}
\mu = \frac{1}{M_z^2} \approx 10.32. \label{scale}
\end{equation}
Finally, performing  the integral in (\ref{coct1}) for the profile 
functions shown in figure \ref{fig:bkg},
we find  that the  counter term is given by,
\begin{equation}
E_{CT}(\tau)=a_1 \left( \frac{1}{\tau}-10.34 \right) + a_2 \log \left( 
\frac{10.34}{\tau} \right), \label{fct1}
\end{equation}
where,
\begin{eqnarray}
a_1 = -3.97886 \times 10^{-4}, \quad
a_2 =  5.76546 \times 10^{-4}. \label{fct2}
\end{eqnarray}
%
The specific parameter values used in the calculation  were,
\begin{eqnarray}
\theta_w = 0.5, \quad
g_u = g_d = 1.0,\quad
\beta = \frac{M_H}{M_Z} = 0.943. 
\label{par}
\end{eqnarray}
The actual background profile are shown in figure \ref{fig:bkg}. The Higgs 
and gauge field profile equations $f(r)$ and $\nu(r)$ were found by 
solving the Nielsen Olesen equation (\ref{int11}), 
(\ref{int12}) for the parameters given above
using numerical relaxation \cite{NUM}. The perturbation 
in the upper component of the Higgs field was taken to be a standard 
gaussian function, so that it is non-zero inside the core of the string 
but decays rapidly outside the core;
\begin{equation}
g(r)=0.1e^{-r^2}.
\end{equation}

As discussed above, we find
two massive bound state solutions in the background on the perturbed 
string. The energies of these bound states are found to be,
\begin{equation}
E^P_{n=0} = 0.99918,\quad
E^P_{n=-1} = 0.0505.
\end{equation}   
In the background of the unperturbed string 
the only massive bound state is a down quark bound state with 
$n=0$ and energy, 
\begin{equation}
E^d_{n=0} = 0.99918.
\end{equation}
This energy is, to this accuracy, identical to that of the massive bound state in the $n=0$ mode
in the perturbed string background.  Hence these two state are naturally  paired 
in the bound state energy contribution. 

The zero modes that appear in the pure  Z-string background naturally pair up with the 
low energy, $n=-1$,  massive bound state in the perturbed string background.

Given this  natural pairing of  the bound states, 
all the other contributions in (\ref{pb}) should vanish for each $n$. That is,
\begin{equation}
\sum_{\alpha=1}^4 \beta_{\infty,P}^{n,\alpha} - \sum_{\pm} \beta_{\infty,d}^{n,\pm} 
- \sum_{\pm} \beta_{\infty,u}^{n,\pm} = 0,
\end{equation}
where the values of the integers $\beta$ are determined by the singularities 
of the functions $X^n(k)$ through (\ref{beg}). This is indeed found to be 
the case. Physically this means that in the spectral flow 
between the unperturbed string and the perturbed string, no continuum states 
drop below threshold energy to become massive bound states.
In accordance with Levinson's theorem, 
  $F^n(k=0)=0$ for all values of $n$, as  there is no difference 
in the number of bound states between the two systems.



\begin{figure}[htp]
\begin{center}
\leavevmode
\epsfxsize=12.0cm
\epsfbox{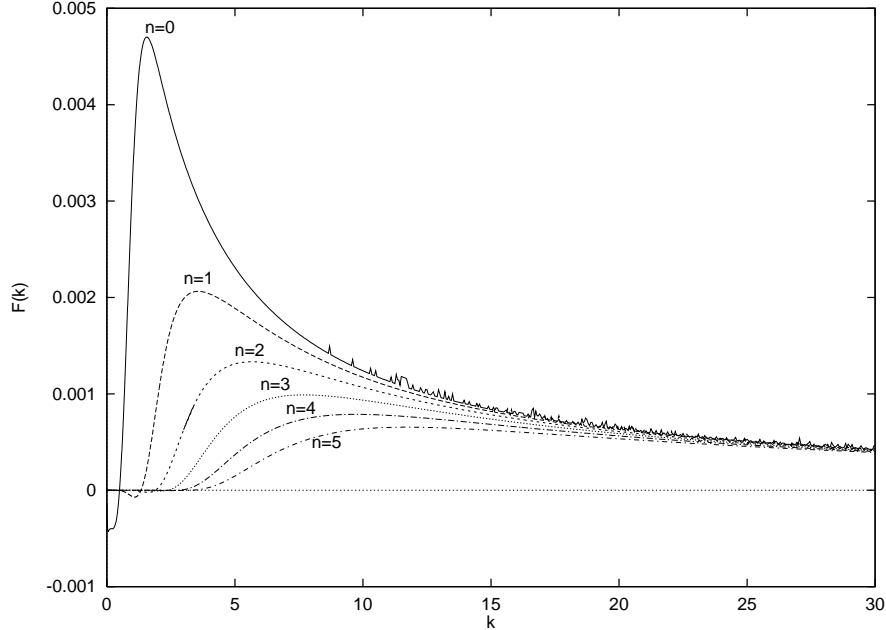}
\caption{The functions $F^n(k)$ for $n=0,1,2,3,4,5$.}
\label{fig:fn05}
\end{center} 
\end{figure}

The functions $F^n(k)$ that define the shift in the energy of the continuum states  
are defined by (\ref{pF}). They have been  found numerically using the formalism developed in
sections 3-6. Figure \ref{fig:fn05} shows plots 
of the functions $F^n(k)$ for $n=0,1,2,3,4,5$, similar behaviour is found for negative $n$.

There are a number of features of the functions $F^n(k)$ worth noting:
Except for a small range around zero momentum, the functions $F^n(k)$ are 
positive for all values of $n$. All other factors comprising the integrand 
for the continuum contribution for each mode are positive 
definite.  However there is an overall minus sign, so each mode 
contributes a negative definite amount to the regularised fermionic energy.

The values of $F^n(k=0)$ are indeed zero for all $n$
(closer inspection of the behaviour of 
$F^0(k)$ for small $k$ reveals the function does indeed vanish).

The functions $F^n(k)$ are {\it enveloped} by the functions with smaller 
absolute mode number.   As  all the other factors in the integrand for the energy contribution 
are independent of $n$, this  implies that as the absolute value of the mode number 
increases the contribution made by the modes to the regularised continuum 
energy  decrease. 

In units of the Higgs VEV,  $F^n(k) \sim 0$ for 
$k$ less than about the absolute value of $n$. This behaviour was noted 
in the case of the core model and is the key to 
the convergence with $n$. Physically, as the  spinors with high angular momentum
are small within the 
core of the string for this range of momenta, they do not couple 
strongly to the perturbation.

The functions $F^n(k)$ die away as $k$ becomes large. Numerical fitting 
to a function of the form $\frac{a_1}{k}+\frac{a_2}{k^2}+ \cdots$ suggests a $\frac{1}{k}$ fall off. 

We can now perform the integrals in 
(\ref{com2}) for specific values of the cut-off $\tau$ and obtain 
the regularised fermionic energy. Numerically, the most intensive part of the calculation  
is the production of the functions $F^n(k)$. These  functions were tabulated 
 for  $n$ between $-20$ and $20$. 

\begin{figure}[htp]
\begin{center}
\leavevmode
\epsfxsize=7.5cm
\epsfbox{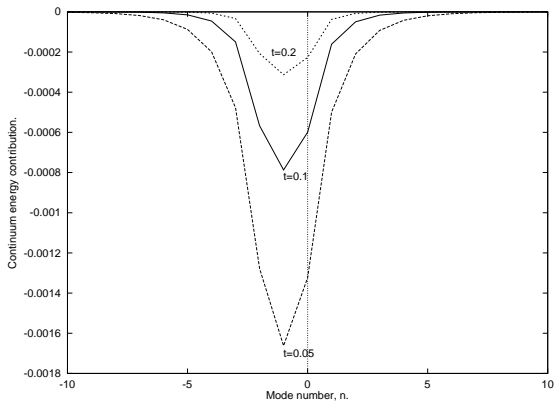}
\caption{A plot of $\delta E^n_{cont}(\tau)$ as a function of $n$ for $\tau=0.2,0.1,0.05$.}
\label{fig:ncoI}
\end{center} 
\end{figure}

Figure \ref{fig:ncoI} shows the continuum contribution to the fermionic energy 
made by each angular mode for $n$ between $-10$ and $10$ and for 
three values of the regulator $\tau$. We see  that the contribution made by each mode
rapidly decreases as the absolute value of the mode number increases,  this implies rapid convergence 
of the the sum over $n$. However, as $\tau \rightarrow 0$,
the contributions from the low $n$  modes can be seen to get larger,   the fall off as 
$\vert n\vert$ gets larger is slower, and  more terms 
must be included in the sum over $n$, as discussed in section 8.4.



We find the regularised fermionic energy for a given value of $\tau$ 
by adding together the contributions made by the bound 
states and the continuum in each mode and then summing over a 
sufficient number of mode to ensure convergence.
The renormalised fermionic energy is then obtained by subtracting the counter term.


There are several sources of numerical error in this calculation. These arise due to
truncation of the sum over angular modes, 
truncation of the momentum integral,
finite tabulation density of the functions 
$F^n(k)$ and numerical errors in the values of these functions.

\begin{figure}[htp]
\begin{center}
\leavevmode
\epsfxsize=7.5cm
\epsfbox{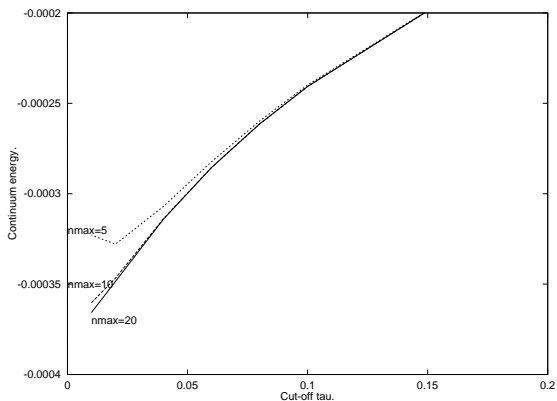}
\caption{Plots of $ \tau \delta E_{cont} (\tau)$ as a function of $\tau$ for different truncations in the sum over the mode number $n$ .}
\label{fig:nbit}
\end{center} 
\end{figure}

The first two errors limit how small the cut off $\tau$ can be taken.
Figure \ref{fig:nbit} shows the regularised continuum energy scaled by the cut off as 
a function of the cut off for three different values of the maximum absolute 
value of the mode number taken in the sum. As 
the number of modes included in the sum decreases, the continuum contribution 
to the regularised fermionic energy becomes less negative. As the contribution made by each mode 
to the regularised energy is negative definite, the truncation in the sum over $n$ induces 
a systematic error in the total regularised fermionic energy. 
 As the number of modes included increases,  $\tau$ can 
be reduced further  before significant errors arise. 

As discussed in section 8.4, the combination of the regulating factor, $e^{-tk^2}$,
and the fact that the functions $F^n(k)$  are small for $ k < n $, leads to 
an exponential suppression of the contribution from high angular modes  by a factor of the
form, $e^{- \tau n^2}$.
For a maximum mode number of $20$ and  
a suppression factor of order $tol$, the smallest value of $\tau$ 
which can be used with confidence is given by,
\begin{equation}
\tau \approx \frac{ \log (tol)}{-400}. \label{yi3}
\end{equation}

To  quantify the errors arising from the intergal over the momentum 
we varied the interpolation procedure and the tabulation density.
Second and third order polynomials were used to interpolate between the tabulated
points and  a second order polynomial fit was made to  only half the data points. 
The error was then estimated from the maximum absolute difference between these three methods. 
Again the error increases as the cut-off is taken away.

The tabulation density used, the truncation $k_{max}=50$ and the truncation $n_{max}=20$  all lead
to errors of order $10^{-6}$ at $\tau \approx 0.04$.  For smaller $\tau$ values the errors grow
rapidly.


Using a quadratic fit to the nodes with $\tau \geq 0.04$ and extrapolating to $\tau=0$, we find,
\begin{equation}
\delta E_{fermion} = -0.0033 \pm 0.00002,
\end{equation}
where the error has been estimated by fitting to fewer nodes.

We have an explicit calculation of the difference between the fermionic ground 
state energies of the perturbed string and the unperturbed string.
This energy difference is negative, as  conjectured  by Naculich \cite{nac}.
 The filled Dirac sea 
further destabilises the string with respect to a perturbation  
of this type. 


\bigskip
\noindent{\bf 9.5 Scaling With Perturbation Size}

Repeating this calculation for a perturbation that is five times larger than the previous one,
we find the same number of bound states and that the functions  $F^n(k)$  have the same characteristic 
features but larger typical values. The 
typical increase in these  values is of the same order as the increase in the 
perturbation squared (i.e. $\sim 25$).

\begin{figure}[htp]
\begin{center}
\leavevmode
\epsfxsize=7.5cm
\epsfbox{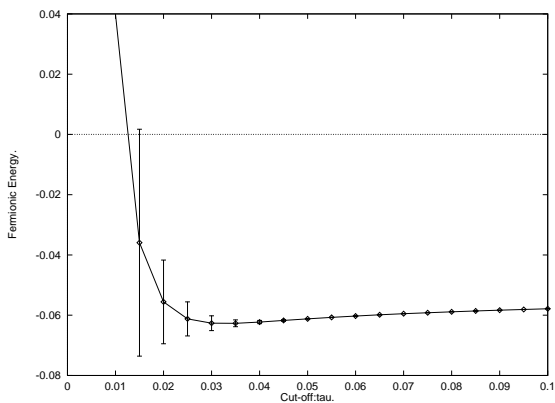}
\caption{A plot of the renormalised fermionic energy $\delta E_{ren}(\tau)$ for
the larger  perturbation.}
\label{fig:near2}
\end{center} 
\end{figure}

Figure \ref{fig:near2} shows a plot of the renormalised energy at a function 
of the cut-off. The extrapolation to obtain the value of the 
fermionic energy was again 
carried using a quadratic fit. The renormalised fermionic 
energy is found to be, 
\begin{equation}
\delta E_{fermion} = -0.0667 \pm 0.0002.
\end{equation}

Again the fermionic energy can be seen to be negative and hence does not 
stabilize the Z-string. 

Comparing the magnitudes of the two fermionic energy shifts, we see that the effect
has increased by roughly  the same order as the increase in the 
perturbation squared (i.e. $\sim 25$).   This is consistent with the suggestions of
Naculich \cite{nac}.

We have calculated the renormalised fermionic energy difference between the 
perturbed and pure Z-string.
 In both cases the fermionic energy is found to be negative and hence has a 
destabilizing effect on the string.  Our values appear to scale roughly with
the square of the magnitude of the perturbation.

\bigskip
\noindent{\bf 9.6 Populating Positive Energy States}

Having calculated the energy shift of the fermionic vacuum,
we can now consider the  effects of populating positive energy bound states.
The inclusion of positive energy bound states is fairly straightforward,
the positive energy states are filled up to  
some given value of the momentum  along the length of the string, $\tilde k$. 
As the perturbation increases, the energy of these states tends to increase and the extra  
contribution to the fermionic energy shift will be positive. At some value of $\tilde k$ 
the contributions from the fermionic ground state and the 
populated positive energy bound states cancel. In this 
situation the fermionic energy has a neutral effect on the stability of 
the string. For higher values of $\tilde k$ the overall energy shift is positive
and the fermions enhance the stability of the string.


It was shown in section 3  that the fermionic contribution to the 
energy of a field configuration is given by,
\begin{eqnarray*}
\sum_r \left( -E_r + n_r E_r \right),    
\end{eqnarray*}
where the sum is over all the positive energy states  and $n_r$ is the 
occupation number of the positive energy states. 
If an adiabatic perturbation were turned on then the change in the fermionic energy is 
given by,
\begin{eqnarray*}
\delta E_{ferm} = \delta E_{sea} + \delta E_{pos},   
\end{eqnarray*}
where $ \delta E_{sea} $ is the sum of the shifts in the negative energy 
states computed above,  and $\delta E_{pos}$ is the shift in the energy of the 
occupied positive energy states. For an adiabatic change, the occupation numbers of the
states remain constant and we have, 
\begin{eqnarray*}
\delta E_{pos} = \sum_r  n_r \left(E_r^\prime -  E_r \right),
\end{eqnarray*}
where $E_r^\prime$ is the shifted value of $E_r$ when the perturbation 
is switched on. In the spectral flow picture these state flow into one 
another.

In the case of the string the states  of interest are the 
zero modes in the unperturbed string background and the low energy 
massive bound state which the up and down quark zero modes combine to 
become when the perturbation is turned on. These
are the only  states between  which there is a significant 
energy shift. 

\begin{figure}[htp]
\begin{center}
\leavevmode
\epsfxsize=10.0cm
\epsfbox{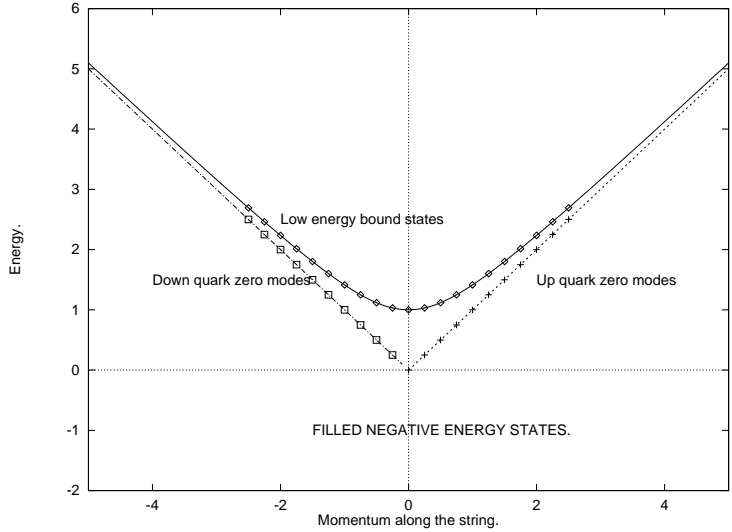}
\caption{A typical scenario illustrating the population of positive energy states.}
\label{fig:current1}
\end{center} 
\end{figure}

We start in the pure Z-string background, with
 up quark zero mode states with momentum in the range $[0 , \tilde k ]$ filled, 
 and down quark zero mode states with momentum in the 
range $[-\tilde k, 0 ]$ filled. 
In the perturbed string this gives
low, positive energy bound states with 
an absolute momentum up to $\tilde k$ filled.
Figure \ref{fig:current1} illustrates the 
population of positive states for the perturbed string (low energy massive 
bound state) and the unperturbed string (up and down quark zero modes) up to 
a value of $\tilde k = 2.5$.
If the low energy bound state has mass $\delta m$, then  a state with 
momentum $k_z$ along the string  satisfies the mass-shell condition, 
$
E^2 = k_z^2 + \delta m^2. \label{cur1}
$
 The zero modes move at the speed of light along the length of the string and 
therefore their energy, $E$, satisfies,
$
E = k_z, \label{cur2}
$ 
 as illustrated by the $45^o$ zero mode lines in 
figure \ref{fig:current1}. 

Discretising the momentum along the string as in section 5, we have,
\begin{equation}
\delta E_{pos} = \sum_{n=-n_{max}}^{n_{max}} \left[ \sqrt{k_n^2+\delta m^2}
- \mid k_n \mid \right],\label{cu3} 
\end{equation}
where, 
$
k_n = \frac{2 \pi n}{L},
$ 
and $L$ is the length of the cylindrical box we are considering.
Taking $L$ to infinity and defining the maximum momentum $\tilde{k}$ by,
$\tilde k = 2 \pi n_{max}/L$,
we obtain,
$$
\delta E_{pos} = {1\over\pi} \int_{0}^{\tilde{k}} dk \bigl( 
\sqrt{k^2+\delta m^2} - \vert k \vert \bigr)
$$
$$
 = {\delta  m^2 \over 2\pi} \biggl[ 
\log \bigl({\tilde{k}\over\delta m} + \sqrt{1+{\tilde{k}^2\over\delta m^2}}\bigr)
 + {\tilde{k}\over\delta m} \bigl(\sqrt{1+{\tilde{k}^2\over\delta m^2}}-{\tilde{k}\over\delta m}\bigr)
\biggr].\label{cu5}
$$

$\delta E_{pos} $ is positive as expected, thus populated positive energy states help to
stabilise the string.

There is also a shift in the energy of any massive bound states when the string is perturbed. 
The energy shift for these states is simply found by considering the  (possibly fictitious)
zero mode as a reference and  taking the difference of \ref{cu5} calculated with
the initial and final masses.  

It may be possible to introduce a sufficient population of positive energy bound states
and zero modes to stabilize the string against the negative curvature induced by the Dirac sea and the 
bosonic instabilities. This question is considered in ref.\cite{Us2}.

\bigskip
\bigskip
\noindent{\bf 10. Conclusions}

\bigskip

We have systematically calculated the order $\hbar$ fermionic energy shift when an electroweak
string is perturbed. This calculation is the first to take into account the effects of the
Dirac Sea. Our approach allows the infinite volume limit to be recovered analytically. We have
developed a state counting procedure that correctly counts states flowing from the continuum to 
bound modes and allows the fermionic energy to be renormalised. The momentum shift functions we
evaluate satisfy Levinson's theorem. Convergence of our numerical sums and the finiteness of the 
renormalised energy shifts have both been demonstrated.

The Dirac Sea energy shift when a pure Z-string is perturbed is found to be negative and 
scales roughly as the size of the perturbation squared. The Dirac Sea thus provides a term with 
negative curvature to the effective potential and has a destabilising effect on the string.

However, populated positive energy states have their energy increased by the perturbation 
and hence have a stabilising effect on the string. The total fermionic effect may vanish
if there is a sufficient population of positive energy bound states on the string. A larger
population of bound states leads to the fermions having a stabilising effect  overall.
This raises the facinating possibility that electroweak strings may be stable if they carry
a fermionic current.

\bigskip
\bigskip
\noindent{\bf Acknowledgements}

\bigskip

The authors would like to thank G. Shore, S. Hands and N. Dorey for useful discussions. WBP would like
to thank A.-C.Davis for useful discussions. MG would like to
thank S.Naculich for useful discussions and acknowledge PPARC for a research studentship.



\bigskip
\bigskip
\noindent{\bf Appendix A: Discretisation Of The Continuum}

\bigskip

 To carry out the  sum over the continuum states we first impose conditions on the spinor profile 
functions on the surface of a cylinder of radius $R_o$ centred on the core of the string. The conditions 
used select a discrete subset of the continuum states and we  trace over this discrete set of states. 
The full continuum is recovered as $ R_o \rightarrow \infty $. This can be done analytically, so that
the  actual trace over the continuum can then be done in the physical, infinite volume limit.
The use of the discretisation condition is merely  an intermediate step in setting up the 
sum over the continuum states, it  is critical as it ensures that the correct 
density of states is used.

To discretise the continuum we demand that 
the $1st$ and $3rd$ components of the spinors vanish at some large radial 
distance $r=R_o$. The physical answer clearly should not depend in anyway 
upon the discretisation condition used and alternatives could have 
been used. The critical point  is that as $R_o$ is taken to infinity, the full continuum 
should be recovered, so all states are included in the sum. That this is true 
for the condition used here will be become clear shortly.

The Dirac equation in 
the Z-string background has two regular short distant solutions and 
two irregular solutions. 
Denoting the regular short distance spinors by $\vec{S}_i$ where $i=1,2$ then 
each of these will asymptotically match onto a unique linear combination 
of the asymptotic solutions given in (\ref{1.46}), 
\begin{eqnarray*}
\vec{S}_i \rightarrow 
\left ( \begin{array}{cccc}
J_{\mid \nu \mid} (\kappa) & 0 & N_{\mid \nu \mid} (\kappa) & 0 \\ 
\mp \frac{E}{k} J_{\mid \nu \mid \pm 1} (\kappa) &
\mp \frac{g}{k} J_{\mid \nu \mid \pm 1} (\kappa) &
\mp \frac{E}{k} N_{\mid \nu \mid \pm 1} (\kappa) &
\mp \frac{g}{k} N_{\mid \nu \mid \pm 1} (\kappa) \\
0 & J_{\mid \nu \mid} (\kappa) & 0 & N_{\nu} (\kappa) \\
\pm \frac{g}{k} J_{\mid \nu \mid \pm 1} (\kappa) &
\pm \frac{E}{k} J_{\mid \nu \mid \pm 1} (\kappa) &
\pm \frac{g}{k} N_{\mid \nu \mid \pm 1} (\kappa) &
\pm \frac{E}{k} N_{\mid \nu \mid \pm 1} (\kappa) 
\end{array}
\right)
\left ( \begin{array}{c}
a_{i,1} \\ a_{i,2} \\ a_{i,3} \\ a_{i,4}
\end{array}
\right),
\end{eqnarray*}
where, $\kappa=KR_0$, the upper sign corresponds to the case $\mid \nu \mid = \nu $ and 
the lower sign $\mid \nu \mid = -\nu $.

By taking an arbitrary linear combination of $\vec{S}_1$ and $\vec{S}_2$ and
setting the $1st$ and $3rd$ components of the spinor to zero at $r=R_o$, 
we obtain the constraints,
\begin{eqnarray*}
 \alpha ( J_{\mid \nu \mid} (k R_o) a_{1,1} + N_{\mid \nu \mid} (k R_o) 
a_{1,3} ) + \beta ( J_{\mid \nu \mid} (k R_o) a_{2,1} + 
N_{\mid \nu \mid} (k R_o) a_{2,3} ) = 0, 
\end{eqnarray*}
\begin{eqnarray*}
 \alpha ( J_{\mid \nu \mid} (k R_o) a_{1,2} + N_{\mid \nu \mid} (k R_o) 
a_{1,4} ) + \beta ( J_{\mid \nu \mid} (k R_o) a_{2,2} + 
N_{\mid \nu \mid} (k R_o) a_{2,4} ) = 0,
\end{eqnarray*}
from which it is easily  seen that,
\begin{equation}
\frac{J_{\mid \nu \mid} (k R_o)}{N_{\mid \nu \mid} (k R_o)} = 
-\frac{(\alpha a_{1,3} + \beta a_{2,3})}{(\alpha a_{1,1} + \beta a_{2,1})}
=
-\frac{(\alpha a_{1,4} + \beta a_{2,4})}{(\alpha a_{1,2} + \beta a_{2,2})}.
\label{e7}
\end{equation} 

Only the ratio $\gamma = \frac{\alpha}{\beta} $ is of interest and it 
is straightforward to derive the following quadratic equation for $\gamma$,
\begin{equation}
 A \gamma^2 + B \gamma + C = 0 ,\label{e8}
\end{equation} 
where,
\begin{equation}
 A = a_{1,3} a_{1,2} - a_{1,1} a_{1,4}, \label{e8.01}
\end{equation}
\begin{equation} 
 B = a_{1,3} a_{2,2} + a_{2,3} a_{1,2} - a_{1,1} a_{2,4} - a_{2,1} a_{1,4} ,
\label{e8.02}
\end{equation}
\begin{equation} 
  C = a_{2,3} a_{2,2} - a_{2,1} a_{2,4}. \label{e8.1}
\end{equation}  
The solution to (\ref{e8}) can be substituted into (\ref{e7}) to produce 
two solutions for each particle,
\begin{equation}
 \frac{J_{\mid \nu \mid}(k R_o)}{N_{\mid \nu \mid}(k R_o)} = -X^\pm (k),
\end{equation}  
where,
\begin{equation}
 X^\pm (k) = \frac{
               (-B \pm \sqrt{B^2-4AC}) a_{1,3} + 2 A a_{2,3} }
               {
               (-B \pm \sqrt{B^2-4AC}) a_{1,1} + 2 A a_{2,1} }.\label{e10}
\end{equation}  

The left hand side of the discretisation condition
is approximately $\cot(kR_o)$, thus  for finite $R_o$ 
there a discrete set of 
string states $\{ k_i \} $ with $ i \in \{ 1,2,3,......\}$ and spacing 
proportional to $ \frac{1}{R_o} $. 
 The density of the solutions becomes infinite as 
$ R_o \rightarrow \infty $ and the full continuum is recovered.

It is possible to 
calculate the solutions of (\ref{e9}) numerically, for a fixed 
value of $R_o$ and then compute the fermionic trace. The process can then be repeated  
for different values of $R_o$ and the limit   $ R_o \rightarrow \infty $ taken 
numerically \cite{Di1}. This is not the approach used here, instead the 
limit  will be taken analytically. This saves the computation 
being repeated for different values of $R_o$.

For momentum $k_i$ such that $k_i R_o \gg 1 $, we can take the leading order asymptotic behaviour of the Bessel 
functions and write the  discretisation condition as,
\begin{equation}
 \cot (k_i R_o - \theta_\nu ) = - X(k_i),\label{e11} 
\end{equation}
where,
\begin{eqnarray*}
 \theta_\nu = \frac{\pi \nu}{2} + \frac{\pi}{4}.
\end{eqnarray*}
This can be inverted to give,
\begin{equation}
 k_i R_o = m(i) \pi + \theta_\nu + \cot^{-1} ( - X(k_i)) \mid_p, \label{e12}
\end{equation}
where $\mid_p$ denotes the principle range of the inverse cotangent, 
throughout this is taken to be $ [0,\pi]$.
$m(i)$ is some integer 
valued function which specifies which branch of the cot function the 
solution lies on.

It is found that $ X(k) $ is singular for certain values of $k$,
the physical significance of such singularities is discussed in sections 8 and 9. 
A singularity in $ X(k) $ corresponds to a discontinuity of $\pi$ in the function 
$\cot^{-1} ( - X(k)) \mid_p$.
As we take the limit the $R_o \rightarrow \infty$,  
the discrete set of solutions, $\{ k_i \}$,  tends towards a continuous spectrum. 
This continuum spectrum, as its name implies, is continuous in the momentum, thus the difference between 
consecutive elements of the set  $\{ k_i \}$ should tend to zero as 
the limit is taken. The requirement of continuity of the spectrum 
and the existence of singularities in $X(k)$ imply that some extra factor must be included. This extra
function, $m(i)$, must jump in integer steps around the singularities of  $X(k)$ to 
ensure that the spectrum defined by $\{ k_i \}$ is indeed continuous as 
$R_o$ is taken to $\infty$. 

It is convenient to write,
\begin{eqnarray*}
 m(i)=i+\beta (i),
\end{eqnarray*}
as this allows an evenly spaced free momentum $k^o_i$ to 
be defined,
\begin{eqnarray*}
k^o_i R_o = i \pi.
\end{eqnarray*}
The function $\beta (i)$ will be combined with the $\cot^{-1} (- X^\pm (k_i)) \mid_p $ term 
to define a continuous function.

In regions where $X(k)$ is finite, we find one solution per branch of the 
cot function, thus $\beta(i)$ is a constant.
However, if $X(k)$ is singular, there can be branches of the cot function
with no solutions or two solutions (fig.\ref{fig:tp2}). 
We denote these singularities as Type 1 and Type 2 respectively.

Each branch of the function 
$\frac{J_{\mid \nu \mid} (k R_o)}{N_{\mid \nu \mid (k R_o)}}$ 
is bounded by two consecutive roots of 
$ N_{\mid \nu \mid} (k R_o) $. Each solution of the discretisation 
condition lies along one of these  branches and is thus bounded by two consecutive 
roots of $ N_{\mid \nu \mid} (k R_o) $. Denoting the $i th$ root of 
$N_\nu (k R_o) $ by $\sigma_i^\nu$, the $ith$ solution of 
(\ref{e9}) will satisfy the inequality, 
\begin{equation}
 \sigma_{i+\delta}^\nu < k_i R_o < \sigma_{i+1+\delta}^\nu, \label{e14}
\end{equation}
where $\delta$ is an integer determined by the 
singularity structure of $ X(k) $. 

If $X^\pm(k)$ has no singularities, it is easy to see that the 
$ith$ string state is bounded by the $ith$ and the $(i+1)th$ 
roots of $N_\nu (k R_o) $ giving  $\delta = 0 $. A Type 1 singularity 
procludes a  solution along one branch, thus
above the singularity  the $ith$ solution is 
bounded by the $(i+1)th$ and $(i+2)th$ roots of $N_\nu (k R_o) $. 
In this case $\delta=1$.
Conversely, a Type 2 singularity gives 
two solutions on one  branch. Thus above the singularity the $ith$ solution is bounded 
by the $(i-1)th$ and the $ith$ roots of $N_\nu (k R_o) $, giving $\delta = -1$.

By counting the singularities of each type, we can determine the value of $\delta$ at momenta greater than that of the
highest momentum  singularity of $X(k)$,
\begin{eqnarray*}
\delta = ( \mbox{ No. Type 1 Singularities} - 
             \mbox{ No. Type 2 Singularities} ).
\end{eqnarray*}     
The asymptotic value of $\beta$, $\beta_\infty$, can then be set such that
the solutions of the discretisation condition fall in the correct range.                
To leading order, the  expansion for the large zeros of a Bessel function gives,
\begin{eqnarray*}
\sigma_m^\nu =
\left\{ \begin{array}{ll}
         m\pi+\theta_\nu-\pi & \nu \geq 0 \\
         m\pi+\theta_\nu-\pi+int[-\nu] & \nu < 0
       \end{array}
\right.
\end{eqnarray*}
To place the high momentum solutions on the correct branches of the cot function
we therefore set,
\begin{equation}
\beta_\infty=\delta -1.\label{beg}
\end{equation}

Having fixed the asymptotic value of $\beta$, we can now consider the 
effect of singularities in $X(k)$. Physically the 
occurrence of such singularities changes the number of continuum states which 
exist. It turns out that the singularities in the functions $X^\pm (k)$ 
are intimately linked to the bound state spectrum and they ensure that the 
total number of fermion states remains the same. The issue is covered in 
detail in section 8, where a specific model is  used 
to demonstrate these points explicitly.

Consider first a Type 1 singularity.
Denoting the the last solution of the discretisation condition before 
the singularity 
by $ k_{i_s^-} $ and the first solution above the singularity by 
$ k_{i_s^+} $, in the case of a Type 1 singularity we see (fig.\ref{fig:tp2}) that, 
\begin{equation}
\cot^{-1} (-X(k_{i_{s^-}})) \mid_p \sim \pi, \label{e23}
\end{equation}
and,
\begin{equation}
\cot^{-1} (-X(k_{i_{s^+}})) \mid_p \sim 0.\label{e24}
\end{equation} 
Using $ i_{{s^+}}=i_{{s^-}}+1 $ and the discretisation condition (\ref{e12}),  we see that,
\begin{equation}
( k_{i_{s^+}} - k_{i_{s^-}} ) R_o = 
  O( \frac{1}{R_o}) + \beta ( i_{i_{s^+}} ) \pi 
   - \beta ( i_{i_{s^-}} ) \pi.\label{e25}
\end{equation} 
As $R_o$ becomes large we require that the momentum shifts are continuous, i.e.  
$( k_{i_{s^+}} - k_{i_{s^-}} ) \sim \frac{\pi}{R_o} $. Thus we have a jump in $\beta$; 
\begin{equation}
\beta ( i_{{s^+}}) = \beta ( i_{{s^-}}) + 1 .\label{e26}
\end{equation}
This is precisely the jump expected due to a Type 1 singularity {\it skipping} a branch of
the cot function. The singularity increases the  value of $\delta$ by one, thus we expect $\beta$ to increase
by one. This jump ensures continuity of the discretisation condition about a Type 1 
singularity. If we start at large momentum and work towards small momentum,  on passing the Type 1 singularity 
the inverse cotangent jumps upwards by $\pi$, but at exactly the same 
point the function $\beta(i) $ jumps down 
by $\pi$: hence there is no overall jump in  the function and continuity 
across the singularity is ensured.

A similar analysis for a Type 2 singularity gives,
\begin{equation}
\beta ( i_{{s^+}}) = \beta ( i_{{s^-}}) - 1 .\label{e27}
\end{equation} 
Again this jump is exactly that expected from a Type 2 singularity giving two solutions on one
branch of the cot function, it ensures continuity of the discretisation condition around a Type 2 
singularity.

It can be seen from (\ref{e26}) and (\ref{e27}) that  
the physical requirement of continuity in the spectrum arises naturally in this formulation.
 
Having fixed the discretisation condition uniquely for all values of the 
momentum the next step is to use the discretisation condition to formulate 
the regularised trace over the continuum states defined by (\ref{e6}). The manner 
in which the discretisation condition has been set up  
allows an integral over the continuum states to be defined in terms of 
the scattering data, found by solving the Dirac equation.
 
The discretisation condition can be written uniquely as, 
\begin{equation}
k_i R_o = k_i^o R_o + \Delta^\pm (k_i),
\end{equation} 
where the free momentum is defined by,
\begin{eqnarray*}
k_i^o R_o = i \pi,
\end{eqnarray*}
and,
\begin{equation}
\Delta^\pm (k) = \beta_\infty + \theta_\nu + \cot^{-1} (-X^\pm(k)) \mid_p^{cont},
\label{e28.1}
\end{equation}
where $ i \in \{ 1,2,.....\}$ and $\cot^{-1} (-X^\pm(k)) \mid_p^{cont}$ is continuous and takes the principal
value for $k$ larger than the last singularity  in $X^\pm(k)$. It should 
be noted that for the pure Z-string there are two discretisation 
conditions of this type for each angular momentum mode labelled by $\pm$ in (\ref{e28.1}). These arise from the two 
roots of the quadratic equation (\ref{e8}) which had to be solved in the process of setting up the 
discretisation condition for the Z-string background.

\bigskip
\bigskip
\noindent{\bf Appendix B: Computation of the Z-string Heat Kernel Coefficients}

\bigskip


The purpose of this appendix is to provide details of the calculation  
of the coincidence limit heat kernel coefficients  required for the 
computation of the counter term in the pure Z-string background.

\bigskip
\noindent{\bf $\mbox{Tr}[a_1]$.}
Using the standard recurrence relations we have,  
\begin{equation}
\mbox{Tr}[a_1]=-a(x)\label{act1}
=-4g^2f^2(r)-2iq(\alpha_L-\alpha_R)\partial_i Z_i 
+iq(\alpha_L-\alpha_R) \partial_i Z_j \mbox{ Tr} \sigma^j \sigma^i. \label{act2}
\end{equation}
Using the standard anti-commutation relation satisfied 
by the Pauli matrices, 
$\{ \sigma^i,\sigma^j \} = 2 \delta^{ij}$, gives, 
$Tr \sigma^i \sigma^j = 2 \delta^{ij}$,  hence the second and 
third terms in (\ref{act2}) cancel, leaving, 
\begin{equation}
Tr[a_1]=-4g^2 f^2(r).\label{act3}
\end{equation}
At this juncture it is however useful to note that individually the second 
and third terms in (\ref{act2}) vanish because the divergence of the 
gauge field in the string 
background is zero, that is $\partial_i Z_i =0$, which is easily shown by 
direct calculation using (\ref{int8}).

\bigskip
\noindent{\bf $\mbox{Tr}[a_2]$.}
Again using the standard recurrence relations we find,
\begin{equation}
\mbox{Tr}[a_2]=\frac{1}{12}Tr \Gamma_{ij} \Gamma_{ij} + \frac{1}{2} 
\mbox{Tr} a^2(x).\label{act4}
\end{equation}
The first term in (\ref{act4}) is straightforward to compute: $\Gamma_{ij}$ 
is the standard field strength tensor;
\begin{equation}
\Gamma_{ij}=\partial_i \Gamma_j - \partial_j \Gamma_i + [ 
\Gamma_i,\Gamma_j].\label{act5}
\end{equation}
In the case of the pure Z-string 
the third term in (\ref{act5}) is zero because of the diagonal nature of 
$\Gamma_i$, hence, 
\begin{eqnarray}
\Gamma_{ij} = \partial_i \Gamma_j - \partial_j \Gamma_i 
= iq Z_{ij} \left(\begin{array}{cc}
\alpha_R & 0 \\
0 & -\alpha_L 
\end{array}
\right), \label{act6}
\end{eqnarray}
where,
\begin{equation}
Z_{ij}=\partial_i Z_j -\partial_j Z_i. \label{act7}
\end{equation}
This is expected because the Z-string is really a Nielsen-Olesen $ U(1)$ string 
embedded in the $ U(1)_Z $ subgroup generated by the Z gauge bosons.
Clearly $Z_{ij}$ is anti-symmetric and so, 
\begin{equation}
 \Gamma_{ij} \Gamma_{ij} = -2 q^2 Z_{12}^2
\left(\begin{array}{cc}
\alpha_R^2 & 0 \\
0 & \alpha_L^2 
\end{array}
\right).
\end{equation}
Using the specific forms of the gauge fields in the string background 
(\ref{int8}) implies that, 
\begin{equation}
Z_{12}=-\frac{\nu^\prime(r)}{r},
\end{equation}
and therefore, 
\begin{equation}
\mbox{Tr } \Gamma_{ij} \Gamma_{ij} = -4 (\alpha_R^2 + \alpha_L^2) 
\left( \frac{\nu^\prime(r)}{r} \right)^2.\label{act0.0}
\end{equation}

As we only need its trace, only the diagonal elements 
of $ a(x)^2 $ need to be considered. Using  
$ \partial_i Z_i = 0 $, it can be shown that, 
\begin{eqnarray}
\mbox{Tr}a(x)^2 &=& 4g^4f(r)^4-q^2(\alpha_L^2+\alpha_R^2)\mbox{ Tr}
(\sigma^j \sigma^i \partial_i Z_j)^2 \nonumber \\
& & +2g^2 \partial_i (f(r) e^{\pm i\theta}) \partial_i(f(r) e^{\mp i\theta}) 
\nonumber \\
& & +2q^2g^2f(r)^2 Z_i Z_i \nonumber \\
& & \pm 2 iqg^2f(r)Z_j \left( e^{\pm i \theta} \partial_i (f(r) 
e^{\mp i\theta}) 
-e^{\mp i \theta} \partial_i (f(r) e^{\pm i \theta}) \right)
. \label{act9}
\end{eqnarray}
Now the explicit forms of the gauge fields given in (\ref{int8}) can be 
substituted into (\ref{act9}). The computation is straightforward but it 
is useful to give the explicit forms  of some of the individual terms 
which appear in (\ref{act9}),
\begin{equation}
Z_i Z_i = \left( \frac{\nu(r)}{q r} \right)^2,\label{act10}
\end{equation}
\begin{equation}
\sigma^j \sigma^i \partial_i Z_j = 
-i \frac{\nu^\prime (r)}{q r} \sigma^3,
\end{equation}
\begin{equation}
\sigma^i \partial_i (f(r) e^{\pm i \theta}) = 
\left(\begin{array}{cc}
0 & e^{-i\theta} \\
e^{i \theta} & 0 
\end{array}
\right)
e^{\pm i \theta} f^\prime (r) \pm
\left(\begin{array}{cc}
0 & e^{- i \theta} \\
-e^{ i \theta } & 0 
\end{array} \right)
e^{\pm i \theta} \left( \frac{f(r)}{r} \right),
\end{equation}
\begin{equation}
Z_i e^{\pm i \theta} \partial_i (f(r) e^{\mp i \theta} = 
\mp i \frac{\nu(r) f(r)}{q r^2}.\label{act11}
\end{equation}
Using (\ref{act10}) to (\ref{act11}) in (\ref{act9}) gives the required 
expression for 
$\mbox{Tr} \hskip 2pt a(x)^2 $ expressed purely in terms of the Nielsen-Olesen profile 
functions and the parameters in the theory. After a little algebra it can 
be shown that, 
\begin{eqnarray}
\mbox{Tr} a(x)^2 &=& 2 g^4 f(r)^4 + {\over 3}2 (\alpha_R^2+\alpha_L^2) 
\left(\frac{\nu(r)^\prime}{r}\right)^2 \nonumber \\
& & +2 g^2 \left( (f^\prime (r))^2+\left( \frac{f(r)}{r} \right)^2 \right) 
\nonumber \\
& & +2 g^2 \frac{f(r)^2 \nu (r)^2}{r^2} + 4 g^2 \frac{f(r)^2 \nu (r)}{r^2}. 
\label{act12}
\end{eqnarray}
It is then easy to show by direct substitution of (\ref{act0.0}) 
and (\ref{act12}) 
into (\ref{act4}) that the expression for $\mbox{Tr}[a_1]$ is indeed that 
given by (\ref{ct12}).

\end{document}